\documentclass[pra,aps,showpacs,floatfix,groupedaddress,10pt]{revtex4-1}

\usepackage{dsfont}
\usepackage{color}
\usepackage{ifthen}
\usepackage{calrsfs}
\usepackage{amssymb}
\usepackage{amsmath}
\usepackage{ifpdf}
\usepackage{graphicx}
\usepackage{dcolumn}
\usepackage{bm}
\usepackage{braket} 
\usepackage[normalem]{ulem}

\newcommand{\mat}[1]{ \underline{\mathbf{#1}} }
\newcommand{\ID}[0]{\mat{\mathrm{1}}}

\begin{document}
\title{Kinetic-Energy Density-Functional Theory on a Lattice}

\author{Iris Theophilou}
\affiliation{Max Planck Institute for the Structure and Dynamics of Matter and Center for Free Electron Laser Science, Hamburg 22761, Germany}

\email{iris.theophilou@mpsd.mpg.de}
\author{Florian Buchholz}
\affiliation{Max Planck Institute for the Structure and Dynamics of Matter and Center for Free Electron Laser Science, Hamburg 22761, Germany}

\author{F. G. Eich}
\affiliation{Max Planck Institute for the Structure and Dynamics of Matter and Center for Free Electron Laser Science, Hamburg 22761, Germany}
 
\author{Michael Ruggenthaler}
\affiliation{Max Planck Institute for the Structure and Dynamics of Matter and Center for Free Electron Laser Science, Hamburg 22761, Germany}

\author{Angel Rubio}

\affiliation{Max Planck Institute for the Structure and Dynamics of Matter and Center for Free Electron Laser Science, Hamburg 22761, Germany}
\affiliation{Center for Computational Quantum Physics (CCQ), The Flatiron Institute, New York NY 10010}

\begin{abstract}
We present a kinetic-energy density-functional theory and the corresponding kinetic-energy Kohn-Sham (keKS) scheme on a lattice and show that by including more observables explicitly in a density-functional approach already simple approximation strategies lead to very accurate results. Here we promote the kinetic-energy density to a fundamental variable along side the density and show for specific cases (analytically and numerically) that there is a one-to-one correspondence between the external pair of on-site potential and site-dependent hopping and the internal pair of density and kinetic-energy density. Based on this mapping we establish two unknown effective fields, the mean-field exchange-correlation potential and the mean-field exchange-correlation hopping, that force the keKS system to generate the same kinetic-energy density and density as the fully interacting one. We show, by a decomposition based on the equations of motions for the density and the kinetic-energy density, that we can construct simple orbital-dependent functionals that outperform the corresponding exact-exchange Kohn-Sham (KS) approximation of standard density-functional theory. We do so by considering the exact KS and keKS systems and compare the unknown correlation contributions as well as by comparing self-consistent calculations based on the mean-field exchange for the keKS and the exact-exchange for the KS system, respectively.

\end{abstract}

\maketitle


\section{Introduction}
\label{sec:intro}

Density Functional Theory (DFT) has become over the past decades a standard approach to the quantum many-body problem. Its success comes from the fact that it combines low computational cost with a reasonable accuracy which helps to understand and predict experimental results for systems not accessible with wavefunction-based methods. DFT avoids the exponential numerical costs of wavefunction-based methods by reformulating quantum mechanics in terms of the density. The major drawback of DFT is that the exact energy expression of the quantum system  in terms of the density is not available and in practice approximations need to be employed. Already before the rigorous formulation of DFT~\cite{Hohenberg-Kohn} a heuristic method based on the density instead of the wavefunction existed which was called Thomas-Fermi theory~\cite{thomas_1927, Fermi1928}. While this theory proved to be very important for the derivation of fundamental results, e.g., the stability of quantum matter~\cite{Stability_of_matter_thomas_fermi}, in practice it is not very accurate (only in the limit of atoms with arbitrarily high atomic number or for homogeneous systems) and does not provide basic properties such as the shell structure of atoms or the binding of molecules. As it was quickly realized it is the approximation to the kinetic energy expression that prevents Thomas-Fermi density-functional approximations to lead to accurate results. What has made DFT popular for determining properties of complex many-body systems is the Kohn-Sham (KS) construction~\cite{Kohn-Sham}, where instead of modeling the kinetic energy directly in terms of the density an auxiliary non-interacting quantum system is used that has the same density. The kinetic energy of this computationally cheap auxiliary system is then corrected by so-called Hartree-exchange-correlation (Hxc) contributions that incorporate the missing interaction and kinetic energy contributions. Already simple approximations to this unknown expression give reasonably accurate answers. However, it is hard to systematically increase the accuracy of approximations while still keeping the favorable numerical costs~\cite{Burke_DFT_review}. Moreover, it has been shown recently that numerous functionals, although accurate when it comes to total energies, fail to reproduce the true density~\cite{Science_Perdew}. The difficulty in functional construction can be attributed to the fact that the Hxc energy depends very implicitly on the auxiliary KS wave function or even more implicitly on the density. 

\par

There are several other approaches for dealing with the quantum many-body problem that also avoid the many-body wave function, while being less implicit. Green's function techniques can be systematically improved in accuracy by including higher-order Feynmann diagrams but are computationally much more expensive~\cite{Fetter_Walecka,Stefanucci_Van_Leeuwen}. Reduced density-matrix (RDM) functional theories~\cite{mazziotti_review, Review-RDMFT_Katarzyna_Klaas} provide a compromise between accuracy and computational cost. In one-body RDM (1RDM) functional theory~\cite{Review-RDMFT_Katarzyna_Klaas} the kinetic energy is an explicit functional of the 1RDM, thus only the part of the interaction energy needs to be approximated, while in the two-body case~\cite{mazziotti_review} even the interaction is given by an explicit functional. Although the explicit use of wave functions can be avoided in these cases, it is still necessary for the RDM to be representable by a wave function.  However, the so called $N$-representability conditions that guarantee an underlying wave function associated with an RDM are anything but trivial~\cite{Nrep-2RDM, Klyachko_GPC}. Moreover, it is not possible to associate to every RDM an auxiliary system of non-interacting particles that would allow to replace the $N$-representability conditions by a numerically simpler auxiliary wave function, like in the DFT case. The Bogoliubov-Born-Green-Kirkwood-Yvon hierarchy,  where the time propagation of an RDM of certain order is related to the RDM of the next order, suffers from similar $N$-representability issues~\cite{Bonitz}.

\par

There are now several possible ways to remedy the above mentioned deficiencies. For 1RDM theory it is helpful to consider the many-body problem at finite temperature and indefinite numbers of particles. In this case the $N$-representability conditions are relaxed and one can even find a non-interacting auxiliary system that generates the same 1RDM~\cite{baldsiefen_reduced_thermal, RDM_elevated_temperature}. Another possibility is to construct approximate natural orbitals which are eigenfunctions of single-particle Hamiltonians with a local effective-potential \cite{Local_RDMFT}. On the DFT side, besides changing the auxiliary system for the KS construction~\cite{gori-giorgi2009, malet2014exchange,grossi2017fermionic}, a possible way out is to include the kinetic-energy density as a basic functional variable along with the density making the modeling of the interacting energy functional less implicit. 
This implies that an additional auxiliary potential which couples to the kinetic energy density has to be introduced.
A similar approach has recently appeared in a different context, i.e., in thermal DFT~\cite{ThermalDFT_PRL,thermal_DFT}, where the additional auxiliary potential corresponds to a proxy for local temperature variations and couples to the entire energy density, including kinetic and interaction contributions. The concept of local temperature was also introduced in the local thermodynamic ansatz of DFT~\cite{Ghosh_DFT_thermodynamics,Ayers_Parr_Nagy_local_temperature,Nagy_dft_thermodynamics}. Furthermore, is important to note that the kinetic-energy density is already used extensively in DFT, for instance, as an integral part of the so-called meta-GGAs. When treated within the generalized KS framework~\cite{SeidlLevy:96} meta-GGAs lead to a local potential coupling to the kinetic-energy density, which can be interpreted as a position-dependent mass~\cite{EichHellgren:14}.

\par

In this paper we investigate the possibility to include the kinetic energy density as a basic functional variable in DFT alongside the density. The idea is that by promoting the kinetic-energy density to an active functional variable  one can increase the accuracy of density-functional approximations. We investigate this by constructing the exact density functionals of standard DFT and compare them vis-a-vis the combined kinetic-energy density and density functionals of this extended approach we call kinetic-energy density-functional theory (keDFT). In this way we want to assess possible advantages of such an approach when considering strongly correlated systems. Further, we want to consider the quality of possible approximation schemes to keDFT based on a kinetic-energy KS (keKS) construction and test them in practice. As is clear from the extend of the proposed program, this is not possible for real systems. Similar to investigations of the exact functionals in DFT~\cite{Tanja_exact_maps_lattice,2site_Hubbard_Burke,Exact_DFT_functional_2sites} and other extensions of DFT~\cite{Tanja_exact_photon_maps}, we restrict our study to a finite lattice approximation for the Hamiltonian, where the particles are only in specific states/positions. We therefore consider lattice keDFT. In this way we do not only avoid the prohibitively expensive calculation of reference data for realistic interacting many-body systems but also avoid mathematical issues connected to the continuum case, like non-existence of ground-states and non-differentiability of the involved functionals~\cite{coarse_grained_DFT, differentiable_DFT} or having to deal with the kinetic energy operator which is unbound~\cite{DFT_math}. All the operators that appear on the lattice are Hermitian matrices which yield lowest energy eigenstates and exact solutions can be easily calculated contrary to the continuum where one always has to resort to basis set approximations. The drawback of this approach is that we cannot follow the usual Hohenberg-Kohn proof in order to show the existence of the basic mapping of keDFT on a lattice (the kinetic-energy density is non-local on the lattice and the spatially dependent mass, i.e., the new external field, becomes part of the kinetic-energy density). While in the appendix we show a proof for a simple case, we construct the basic maps numerically to provide a basis for lattice keDFT. We also highlight how simple approximations carry over from our model systems to more complex lattice systems and even to the full continuum limit. The results hint at the possibility to treat weakly and strongly correlated systems with the same simple approximation to keDFT. 

\par

The paper is structured as follows: In Sec.~\ref{sec:Setup} we introduce the Hubbard model, define the density and kinetic-energy density on the lattice and highlight for a simple two-site case that the kinetic-energy density is a natural quantity to be reproduced by an extended KS construction. We then introduce the resulting keKS construction assuming the existence of the underlying maps between densities and fields. In Sec.~\ref{sec:Mapping} we discuss these mappings and show how by allowing a spatially dependent mass/hopping a large gauge freedom is introduced. Still we can provide a bijective mapping between densities and fields for specific cases. In Sec.~\ref{sec:Inversion} we then show how we numerically construct the mappings beyond these specific cases and hence find that keDFT on a lattice can be defined also for more general situations. In Sec.~\ref{sec:Results} we then use the constructed mappings to determine the exact correlation expressions for the KS and the keKS construction, respectively. In Sec.~\ref{sec:SCF_Results} we then compare the results of self-consistent calculations for similar approximations for the KS and the keKS systems, respectively. Finally, we conclude in Sec.~\ref{sec:Outlook}.

\section{Formulation of the lattice problem}
\label{sec:Setup}

In the following we consider quantum systems consisting of $N$ fermions (electrons) on a one dimensional lattice of $M$ discrete sites. We assume that these particles can move from site to site only via nearest-neighbor hopping (corresponding to a second-order finite-differencing approximation to the Laplacian) and employ zero boundary conditions for definiteness (the extension to periodic boundary conditions is straightforward). This leads to  
a Hamiltonian of the following type 
\begin{eqnarray}
\label{interacting_Ham}
\hat{H}= -t \sum_{i=1,
\sigma=\uparrow,\downarrow }^{M-1}
(\hat{c}_{i}^{\sigma\dagger}\hat{c}^{\sigma}_{i+1}+h.c.)\nonumber\\
+\sum_{i=1}^M v_i\hat{n}_i
+U\sum_{i=1}^{M}\hat{n}^{\uparrow}_i \hat{n}^{\downarrow}_i.
\end{eqnarray}
The non-local first term corresponds to the kinetic energy. Without loss of generality we can assume that the hopping amplitude obeys $t > 0$. The second term corresponds to a local scalar electrostatic potential $v_i$ acting on the charged particles at site $i$. $U \geq 0$ is the on-site Hubbard interaction between the fermions, which is a reminiscence of the Coulomb interaction. Further, the fermionic creation and annihilation operators obey the anti-commutation relations $\{\hat{c}_{i}^{\sigma\dagger},\hat{c}_{j}^{\sigma'}\}=\delta_{ij}\delta_{\sigma \sigma'}$, where $\sigma$ corresponds to the spin degrees of freedom of the particles, $\hat{n}^{\sigma}_i= \hat{c}_{i}^{\sigma\dagger}\hat{c}^{\sigma}_i$ is the spin density operator and $\hat{n}_i=\hat{n}^{\uparrow}_i+\hat{n}^{\downarrow}_i$ is the density operator that couples to the electrostatic potential. Since we fix the number of particles the potential $v_i$ is physically equivalent to a potential that differs by only a global constant. In the following this arbitrary constant is fixed by requiring
\begin{eqnarray}
\sum_{i=1}^{M} v_i=0.
\label{gauge_v}
\end{eqnarray}
Now, if $\Psi$ is the ground-state wave function of Hamiltonian~\eqref{interacting_Ham} we can associate to every point in space a ground state density $n_i=\langle \Psi|\hat{n}_i|\Psi\rangle$. From the lattice-version of DFT~\cite{Chayes_Ruskai} we know that for every fixed set of parameters $(t,U)$ we find then a bijective mapping between the set of all possible potentials (in the above gauge) to all possible densities for a fixed number of particles. To ease notation we introduce a vector for the density $\mathbf{n} \equiv (n_1,...,n_M)$ and accordingly for the potential $\mathbf{v} \equiv (v_1,...,v_M)$, which allows us to write the underlying mapping as $\mathbf{n} \overset{1:1}{\mapsto} \mathbf{v}$. Accordingly, for the potential of an interacting system ($U >0$) as functional of the density we write $v_{i}[\mathbf{n}]$. We further note that since the total number of particles is fixed to $N$, the density is constrained by $\sum_{i=1}^M n_i =N$. This means that instead of the density at every point one can equivalently use the density differences between sites $\Delta n_i= \langle \Psi |\hat{n}_i-\hat{n}_{i+1}|\Psi\rangle$ to establish the above mapping at fixed number of particles. Similarly, knowing the local potential $v_i$ at every site together with the gauge condition~\eqref{gauge_v} is equivalent to knowing $\Delta v_i=v_i-v_{i+1}$. In certain situations, e.g., for figures, it is more convenient to use the density and potential differences instead of the density and potential.

\par

Clearly a similar mapping also holds for a non-interacting Hamiltonian, i.e., $U=0$. Since it is bijective, we can invert the mapping and find a potential  $\mathbf{v}^{s}$ (where we follow the usual convention and denote the potential of a non-interacting system with an $s$) for a given density $\mathbf{n}$. The non-interacting mapping allows to define $v^{s}_i[\mathbf{n}]$, which in turn leads to
\begin{eqnarray}
\hat{H}^{s}= -t \sum_{i=1,\sigma=\uparrow, \downarrow}^{M-1} (\hat{c}_{i}^{\sigma\dagger}\hat{c}_{i+1}^{\sigma}+h.c.)\nonumber\\
+\sum_{i=1}^M v_i^{s}[\mathbf{n}] \hat{n} _i.
\label{S_Ham}
\end{eqnarray}
By construction this non-interacting Hamiltonian reproduces the prescribed density $\mathbf{n}$ as its ground state. This is not yet the KS construction, since we need to know the target density in advance. Only upon connecting the interacting with the non-interacting system by introducing the Hxc potential
\begin{eqnarray}
v^{Hxc}_{i}[\mathbf{n}] = v^{s}_{i}[\mathbf{n}] - v_{i}[\mathbf{n}],
\label{v_Hxc}
\end{eqnarray}
which can also be defined as a derivative of the corresponding Hxc energy functional (see Sec.~\ref{sec:Results}) with respect to $\mathbf{n}$, we find the non-linear KS equation for a given and fixed external potential $\mathbf{v}$ of the interacting system 
\begin{eqnarray}
\hat{H}^{KS}= -t \sum_{i=1,\sigma=\uparrow, \downarrow}^{M-1} (\hat{c}_{i}^{\sigma\dagger}\hat{c}_{i+1}^{\sigma}+h.c.)\nonumber\\
+\sum_{i=1}^M (v_i + v_i^{Hxc}[\mathbf{n}]) \; \hat{n} _i.
\label{KS_Ham}
\end{eqnarray}
This problem has as the unique solution the non-interacting wave function that generates the density of the interacting problem without knowing it in advance~\cite{existence_TDDFT}. To make this scheme practical, one needs to employ an approximation for the Hxc potential, (where the simplest would be a mean-field ansatz of the form $v_i^{Hxc}[\mathbf{n}] \approx U n_i$). 

\par

As we will see in Sec.~\ref{sec:Results}, the major problem in these approximations is that the kinetic energy density of the KS and the interacting system become dramatically different with an increasing $U$. Here the kinetic-energy density $T_i$ at site $i$ is defined non-locally (because it involves the hopping) with the help of the first off-diagonal of the (spin-summed) 1RDM in site basis representation 
\begin{eqnarray}
T_{i}=-t (\gamma_{i,i+1}+ \gamma_{i+1,i})
\end{eqnarray}
where $\gamma_{i,i+1}$ is given by
\begin{eqnarray}
\gamma_{i,i+1}=\langle \Psi|\hat{\gamma}_{i,i+1}|\Psi\rangle
\textrm{  with  } 
\hat{\gamma}_{i,j}=\sum_{\sigma}
\hat{\gamma}_{i,j}^\sigma,\textrm{ and }
\hat{\gamma}_{i,j}^\sigma=\hat{c}_{i}^{\sigma\dagger}\hat{c}_{j}^{\sigma}.
\end{eqnarray}
By analogy to the continuum case, one can also define the charge current $J_i$ as
\begin{eqnarray}
J_{i}=-it (\gamma_{i,i+1}-\gamma_{i+1,i})
\end{eqnarray}
With no external magnetic field present, i.e., no complex phase of the hopping amplitude, the ground state wave functions are real valued, which implies $\gamma_{i,i+1}=\gamma_{i+1,i}$, leading to zero current.
We note that the current obeys the lattice version of the continuity equation
\begin{eqnarray}
\dot{n}_i =-\mathcal{D}_{-}J_i
\label{continuity}
\end{eqnarray}
in a time-dependent situation, where $\mathcal{D}_{-}J_i = J_i-J_{i-1}$ is the backward derivative of $J_i$.

\par

Clearly, if we could enforce that an auxiliary non-interacting system has the same 1RDM as the interacting one, then also the kinetic-energy densities $\mathbf{T}$ of the two systems would coincide. This suggests to establish a mapping between the interacting 1RDM and non-local potentials, i.e., a $v_{i,j}$ that connects any two sites of the lattice and thus couples directly to the full 1RDM. However, in general this is not possible as has been realized early on in 1RDM functional theory~\cite{Gilbert:75}. A concrete example is the two-site homogeneous Hubbard problem at half filling. In this case, we have $i={1,2}$ and $v_i =0$. So we have a homogeneous density $n_i = 1$ and we can analytically determine all eigenfunctions of the interacting and non-interacting system. Further, in the case of only two sites the full 1RDM is a $2 \times 2$ matrix, where the diagonals are merely $\gamma_{i,i} = n_{i} = 1$ and the off-diagonals are given explicitly by $\gamma_{1,2} = \gamma_{2,1} = \frac{4t}{\sqrt{{(4t)}^2 +U^2}}$. Since the density fixes the potential of the interacting and KS system to be exactly zero, our only freedom is to adopt the non-local potential which is equivalent to just adopting the hopping of the KS system (in this case the non-local potential $v_{1,2} \equiv t$). But since the off-diagonals for the KS system are $\gamma_{1,2}^s = \gamma_{2,1}^s \equiv 1$ irrespective of the hopping amplitude, no non-local KS potential exists which reproduces the interaction 1RDM. This is also true in more general lattice situations as has been shown in, e.g., Ref.~\cite{v-representability_gamma_lattice}. For the 1RDM, two solutions to this problem are known. Either we include temperature and possibly an indefinite number of particles, which introduces off-diagonals that depend on the temperature and the hopping, i.e, the non-local potential~\cite{RDM_elevated_temperature}. We note that for the homogeneous two-site case this can still be solved analytically and verified explicitly. The other possibility is to make the system degenerate such that we can reproduce any density matrix~\cite{RDM_elevated_temperature}. 

\par 

Here, we apply a different strategy. While we cannot force the density matrices to coincide, it is possible to require the kinetic energy densities to be the same. The crucial difference is that we include the coupling in the Hamiltonian in the definition of the quantity to be reproduced by the KS system. For example, in the two-site case we merely need to use an interaction-dependent hopping $t^{ke}=\frac{8t^2}{\sqrt{{(4t)}^2 +U^2}}$. Thus, the auxiliary non-interacting system reproduces now the pair $(\mathbf{n},\mathbf{T})$ of the interacting system. Before we move on, let us note that similarly to the continuum case one could use 1RDM functional theory at zero temperature also on the lattice if one avoids the use of a non-interacting auxiliary system and merely uses functionals based directly on the interacting 1RDM~\cite{Pastor,Mueller_Hubbard}. Note that $N$-representability would need to be enforced in such a scheme.

\par 

Let us now assume that similar to DFT we can establish a bijective mapping 
\begin{eqnarray}
(\mathbf{v}, \mathbf{t}) \overset{1:1}{\mapsto}
(\mathbf{n}, \mathbf{T})
\label{mapping}
\end{eqnarray}
which would allow us to define hopping parameters and potentials that generate a given kinetic-energy density and density, i.e., $t_i[\mathbf{n}, \mathbf{T}]$ and $v_i[\mathbf{n}, \mathbf{T}]$. Specifically we can then consider a non-interacting auxiliary problem that generates a prescribed pair $(\mathbf{n},\mathbf{T})$
\begin{eqnarray}
\hat{H}^{s}= -\sum_{i=1}^{M-1} t_i^{s}[\mathbf{n},\mathbf{T}](\hat{\gamma}_{i,i+1}+h.c.)\nonumber\\
+\sum_{i=1}^M v_i^{s}[\mathbf{n},\mathbf{T}] \hat{n} _i
\label{ke_Ham}
\end{eqnarray}
by its groundstate $\Phi^{ke}$. If we introduce then the according mapping differences similar to Eq.~\eqref{v_Hxc} and denote them by mean-field exchange-correlation (Mxc) 
\begin{align}
v^{Mxc}_{i}[\mathbf{n},\mathbf{T}]& = v^{s}_{i}[\mathbf{n},\mathbf{T}] - v_{i}[\mathbf{n},\mathbf{T}],
\label{v_Mxc}
\\
t^{Mxc}_{i}[\mathbf{n},\mathbf{T}] &= t^{s}_{i}[\mathbf{n},\mathbf{T}] - t_{i}[\mathbf{n},\mathbf{T}],
\label{t_Mxc}
\end{align}
we find the corresponding keKS system
\begin{eqnarray}
\hat{H}^{ke}= -\sum_{i=1,\sigma=\uparrow, \downarrow}^{M-1} & (t_i + t_i^{Mxc}[\mathbf{n},\mathbf{T}])&(\hat{\gamma}_{i,i+1}+h.c.)\nonumber\\
+\sum_{i=1}^M &( v_i + v_i^{Mxc}[\mathbf{n},\mathbf{T}])& \hat{n} _i
\label{keKS_Ham}
\end{eqnarray}
such that
\begin{eqnarray}
T_{i}=-t\langle \Psi|\hat{\gamma}_{i,i+1}|\Psi\rangle+c.c.=\nonumber\\
-t_i^{s}[\mathbf{n},\mathbf{T}]\langle \Phi^{ke}|\hat{\gamma}_{i,i+1}|\Phi^{ke}\rangle+c.c.
\end{eqnarray}
and
\begin{eqnarray}
n_i=\langle\Psi|\hat{n}_i|\Psi\rangle=\langle \Phi^{ke}|\hat{n}_i|\Phi^{ke}\rangle.
\end{eqnarray}
This construction gives rise to the keKS hopping $t^{ke}[\mathbf{t}; \mathbf{n}, \mathbf{T}]$ and the keKS potential $v^{ke}[\mathbf{v}; \mathbf{n}, \mathbf{T}]$. In order to make the scheme practical we now need two approximations: one for the Mxc potential and one for the Mxc hopping. Before discussing possible routes on how to construct approximations and how this could help to more accurately capture strongly-correlated systems, we investigate whether the above proposed mappings do exist.

\section{Generalized mappings from densities to potentials}
\label{sec:Mapping}

\begin{figure}
\includegraphics[width=0.95\columnwidth]{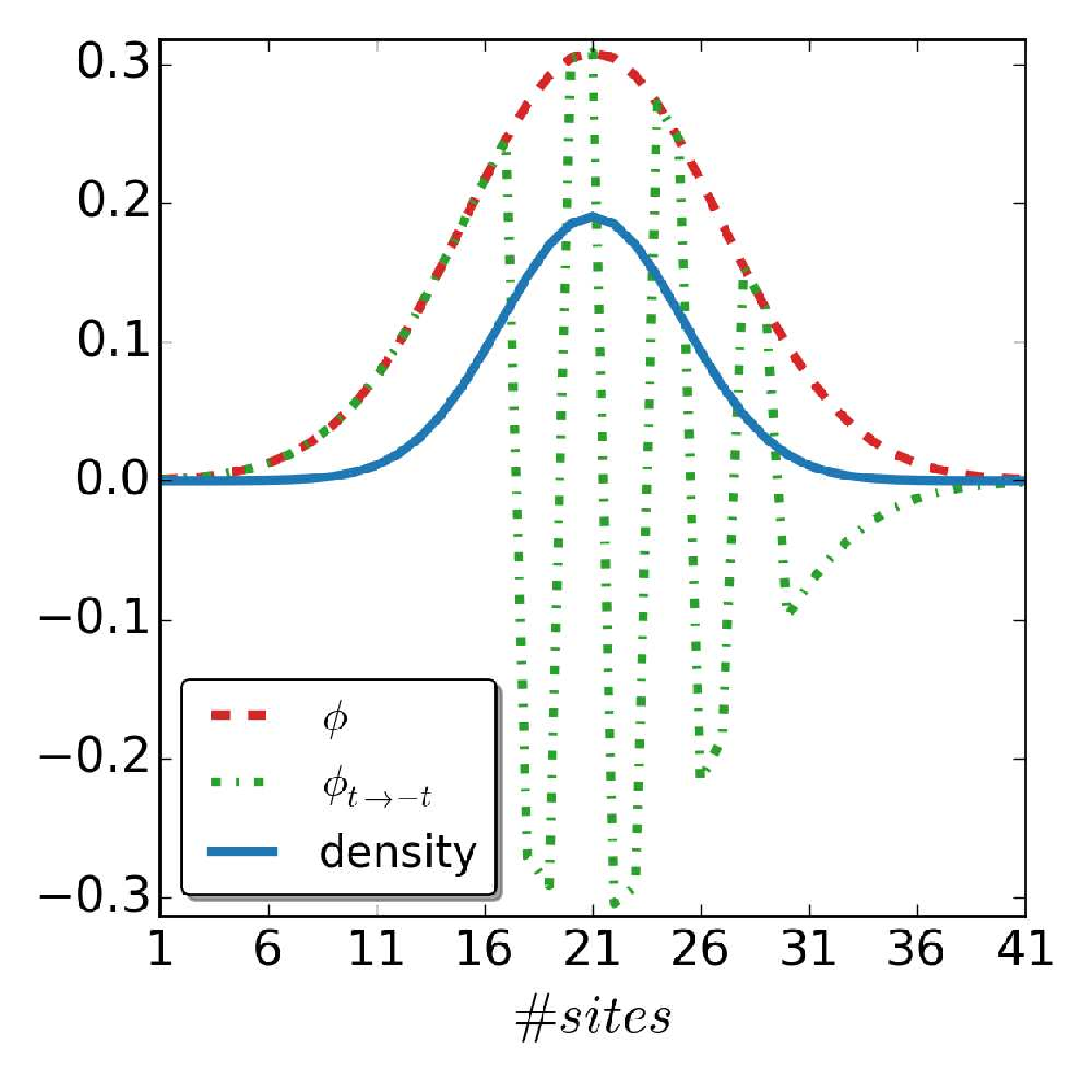}
\caption{The figure shows the doubly occupied orbital $\phi$ that corresponds to a two electron singlet-state of a single particle Hamiltonian with all hopping parameters $t^{s}_{i} = t$ positive and the corresponding one $\phi_{t\rightarrow-t}$ with alternating hopping parameters $\pm t$ from site 17 to 29. Every time we alternate $t$ to $-t$ at site $i$ the orbital $\phi$ changes sign from that site onwards. Since we replace $t$ to $-t$ from site to site, the orbital will recover its original sign after two sites. As one can readily see, the density stays the same in both cases, as a consequence of the sign of $t$ being only a gauge choice. }
\label{gauge-t}
\end{figure}

Similarly to fixing the constant of the local potential, one needs to fix the gauge of the hopping parameter. One of the first things to note is that by letting $t_i$ change from site to site is that we encounter a large equivalence class for the site-dependent hopping parameters. Indeed, we can arbitrarily change the signs of the hopping from $t_i^{s} \rightarrow -t_i^{s}$ without changing the density and the kinetic energy density. However, the wavefunction and also, e.g., the 1RDM, changes. Indeed, changing the sign locally, say at site $i$, will transform the single-particle wavefunction at this site $\phi_{i}$ to $-\phi_{i}$ (see Fig.~\ref{gauge-t} for an example and App.~\ref{app:Gauge} for further details). This leaves the density unchanged, as it is just a sum of the squared absolute values of the single-particle wave functions. Also the kinetic-energy densities stay the same, since the 1RDM switches signs at the same place as the hopping amplitude. As it follows from the discussion above, the sign of $t_i$ is just a gauge choice and we need to fix the gauge in order to establish the sought-after mapping in the non-interacting case. In the following we choose $t^{s}_i >0$.

\par

A further complication that one encounters in establishing the necessary mappings is that the usual Hohenberg-Kohn approach does not work in our lattice case. The reason is that the control fields $t_i$ now become explicitly part of the control object $T_i$. A similar problem is encountered in current-density functional theory, when trying to establish a mapping in terms of the gauge-independent physical charge current \cite{VignaleRasolt:87,Diener:91,Vignale:04,TDCDT_lattice_proof,Ullrich:12}. While in the time-dependent case having the field as part of the control objective is actually an advantage and a general proof has been established~\cite{TDCDT_lattice_proof}, these complications unfortunately prohibit a simple general proof of the existence of the mapping $(\mathbf{n},\mathbf{T}) \mapsto (\mathbf{v},\mathbf{t})$ for the time-independent case. 
However, for specific situations we are able to show that the discussed mapping is possible. The most important one in our context is the case of the two-site Hubbard model (see App.~\ref{app:proof_2_sites} for details). In this case we only have a single potential difference $\Delta v$ and density difference $\Delta n$. So we can simply rescale the auxiliary Hamiltonian and thus prove the existence of the mapping in the non-interacting case by employing the Hohenberg-Kohn results.
A further simple case is two non-interacting particles in a general $M$-site lattice. Here the density fixes the single-particle orbital (doubly occupied) up to a sign and thus for a given $T_i$ only a unique site-dependent hopping $t_i$ is possible. Finally, in the homogeneous case, where the local potential $v_i=0$ and periodic boundary conditions are employed, the density $n_i=\frac{N}{M}$ and the kinetic-energy density of the interacting system will be constant at every site i, $T_i=T$. The matrix elements $\gamma_{i,i+1}$ will also be constant from site to site, $\gamma_{i,i+1}=\gamma$. In this case the mapping is invertible and a unique (up to a sign choice) $t_i=-\frac{T}{2\gamma}$ is associated from site to site. Note that in this case
the KS system and the keKS system yield the same wavefunction and $\frac{t^{ke}}{t}=\frac{T}{T^{KS}}$. This last example, although it only shows the invertibility of the mapping $( \mathbf{v},\mathbf{t}) \mapsto (\mathbf{n}, \mathbf{T})$ at the specific points $t_i=t > 0$ and $v_i=0$, has very important consequences. It allows us in a simple yet exact way to connect the auxiliary keKS system to the interacting system. We will use this later to construct a first approximation to $t^{Mxc}_i$.

\par

To show that the keDFT mapping can also be defined for other, more general cases, we construct in the following the mappings numerically. Afterwards we then make use of the constructed mappings to investigate the properties of the Mxc potentials and the basic functionals, which for the continuum case would be numerically prohibitively expensive. 

\section{Inversion of $(\mathbf{n}, \mathbf{T})$}
\label{sec:Inversion}

Since, as discussed above, it is not straightforward to show that the mapping~\eqref{mapping} is 1-1 in general, we investigate this question numerically.
Therefore we construct sets of densities and kinetic-energy densities $(\mathbf{n}, \mathbf{T})$ by solving the interacting problem specified by the Hamiltonian given in  Eq.~\eqref{interacting_Ham} and for every set we determine the potentials $(\mathbf{v},\mathbf{t})$ of the non-interacting Hamiltonian specified in Eq.~\eqref{keKS_Ham} which yields the target densities $(\mathbf{n},\mathbf{T})$. To determine these potentials, we set up an inversion scheme by using the equations of motion (EOM) for the density and the kinetic-energy density, respectively. These provide not only physical relations that connect the quantities $(\mathbf{v},\mathbf{t})$ with $(\mathbf{n},\mathbf{T})$, but they are also suitable to define correlation potential, as we will explain in the following. 
\par
Note, that in principle the inversion can be done with other techniques, which are used to find the exact local KS potential for a given interacting target density~\cite{RVL_inversion,direct_optimization_inversion, Inversion_review_Wasserman,Staroverov_inversion}. However, it is not clear whether all these techniques can be successfully transfered to the current situation. For instance, in Ref.~ \cite{RVL_inversion} an iteration scheme is introduced that adopts the potential based on the intuition that where the density is too low the potential is made more attractive and where the density is too high it is made less attractive. It is not so clear how to transfer this intuitive procedure to the kinetic energy density $T_i$ which is non-local.

\par

Since the the first order EOM for the density, i.e. the continuity equation~\eqref{continuity}, is trivially satisfied as the current is just zero in the ground state, we consider the second time derivative of the density $\ddot{n}_i$. Since the first time derivative of the kinetic-energy density vanishes for ground-state wave functions, we use again the second-order EOM $\ddot{T}_i$.

As examples, we give here the EOMs for $\ddot{n}^s_1$ and $\ddot{T}^s_1$ for two sites that we use in our numerical inversion scheme:

\begin{eqnarray}
\ddot{n}^s =2{(t^s)}^2\Delta n^s-\Delta v^{s} T^s
\label{eq_mot_n_2sites}
\end{eqnarray}
\begin{eqnarray}
\ddot{T}^s=-\Delta v^s\left(2{(t^s)}^2\Delta n^s-\Delta v^{s} T^s\right)=-\Delta v^s \ddot{n}^s
\label{eq_mot_T_2sites}
\end{eqnarray}
Here we have dropped the site index since everything corresponds to site 1, $\Delta n^s=(n_1-n_2)^s$ is the density difference between the two sites $\Delta v^s=(v_1^s-v_2^s)$ is the local potential difference and $T^s=-2t^s\gamma_{1,2}^s$.  As one can readily see for the 2 site case there is no additional information in the equation for $\ddot{T}^s$ as once $\ddot{n}^s=0$, $\ddot{T}^s$ is also zero. Nevertheless, once we go to more sites $\ddot{T_i}^s$ will also give us new equations. For a detailed discussion of this issue see App.~\ref{app:eqs_motion}.

\par

The inversion scheme we employ is an iterative procedure based on the above introduced EOMs (see Eqs.~\eqref{eq_mot_ni} and \eqref{eq_mot_Ti_2} in App.~\ref{app:eqs_motion} for the general expressions), which provide us with relations between $(\mathbf{\Delta v}^{s}, \mathbf{t}^{s})$ and the target quantities $(\mathbf{n},\mathbf{T})$. The target quantities $(\mathbf{n},\mathbf{T})$ we obtain by finding the ground state of the corresponding interacting Hamiltonian of Eq.~\eqref{interacting_Ham}. We then choose as initial guess for the auxiliary keKS system the values of the interacting system $v_i^{s,0}=v_i$ and $t^{s,0}_i=t$. 
\newline

(a) We solve the auxiliary non-interacting Schr\"odinger equation \eqref{ke_Ham} with the values $v_i^{s,0}$ and $t^{s,0}_i$, 
\begin{eqnarray}
\left (\sum_{i=1}^{M-1} t_i^{s,0}(\hat{\gamma}_{i,i+1}+h.c.)
+\sum_{i=1}^M v_i^{s,0}\hat{n} _i\right)
|\Phi^0\rangle=\epsilon |\Phi^0\rangle.
\label{eq:Scrh_aux}
\end{eqnarray}

(b) We next calculate the density and kinetic-energy density that correspond to the state $|\Phi^0\rangle$, i.e., $n_i^0=\braket{\Phi^0|\hat{n}_i|\Phi^0}$ and $T_i^0=2t_i^{s,0}\langle \Phi^0|\hat{\gamma}_{i,i+1}|\Phi^0\rangle $ as well as the matrix elements $\gamma_{i,j}^0$ that enter the EOMs~\eqref{eq_mot_ni} and \eqref{eq_mot_Ti_2}.\newline

(c) In a last step, we then calculate the variables of the next iteration $v^{s,1}_i$ and $t^{s,1}_i$. The EOM for $\ddot{n_i}=0$ of Eq.~\eqref{eq_mot_ni} provides us with analytic expressions of $v^{s,1}_i$ in terms of the target densities, the hopping amplitudes $t^{0}_{i}$ and reduced density matrix elements $\gamma_{i,j}^0$ of the previous iteration. For calculating the $t^{s,1}_{i}$, we use a numerical solver on all the available EOMs for $\ddot{n}_i$ and $\ddot{T}_i$, with the target kinetic-energy densities, but updated densities $n_i^0$ and $\gamma_{i,j}^0$ from the last iteration and the renewed local potentials $v^{s,1}$. We repeat the steps (a) to (c) until convergence of the calculated fields.

\par

As an example in the 2-site case one can update in every iteration the local potential
\begin{eqnarray}
\Delta v^{s,i}=\frac{{2(t^{s,i-1})}^2\Delta n}{T^{i-1}}
 \label{v1_iter}
\end{eqnarray}
and the hopping parameter
\begin{eqnarray}
 t^{s,i}=\left(\frac{T \Delta v^{s,i-1}}{2\Delta n^{i-1}}\right)^{1/2},
 \label{t1_iter}
\end{eqnarray}
where $\Delta n$ is the target density difference between the two sites and $T$ is the target kinetic-energy density. 

\par

We want to point out that the procedure to update $\mathbf{v}^{s,i}$ and $\mathbf{t}^{s,i}$ is not the only one possible. For example, one could have used instead of the EOMs that we get for $\ddot{n}_i=0$ the ones for $\ddot{J}=0$. Further note that there are always $M-1$ independent equations from $\ddot{n}_i=0$ because of particle number conservation, thus as many as the independent $v_i$ that we have. The number of EOMs that we get for the kinetic energy density $\ddot{T}_i$  is $M-2$ , as we explain in App.~\ref{app:eqs_motion}. The interacting ground state was obtained using the single-site DMRG~\cite{hubig15} routine, implemented in the~\textsc{SyTen} toolkit~\cite{hubig17_2}.

\par

We successfully performed inversions for systems on up to four sites with different total number of electrons for different on-site interaction strengths $U$ and local potentials $\mathbf{v}$. Some representative results are shown in the next section, where we use the constructed mappings to consider the exact keKS system. We also performed successful inversions for the same systems for the interacting problem, i.e., we chose random values $(\mathbf{n}, \mathbf{T})$ and reproduced them with a non-zero Hubbard interaction. This makes the equations involved slightly more complex (and we refrained from showing them here explicitly), but the inversion procedure stays the same. The fact that we could indeed construct a keKS auxiliary system for these cases as well as perform inversions for the interacting problems provides us with indications for the existence of a keKS system for arbitrary number of electrons/sites.

\section{Comparing the exact KS and keKS construction}
\label{sec:Results}

\begin{figure}
\includegraphics[width=0.95\columnwidth]{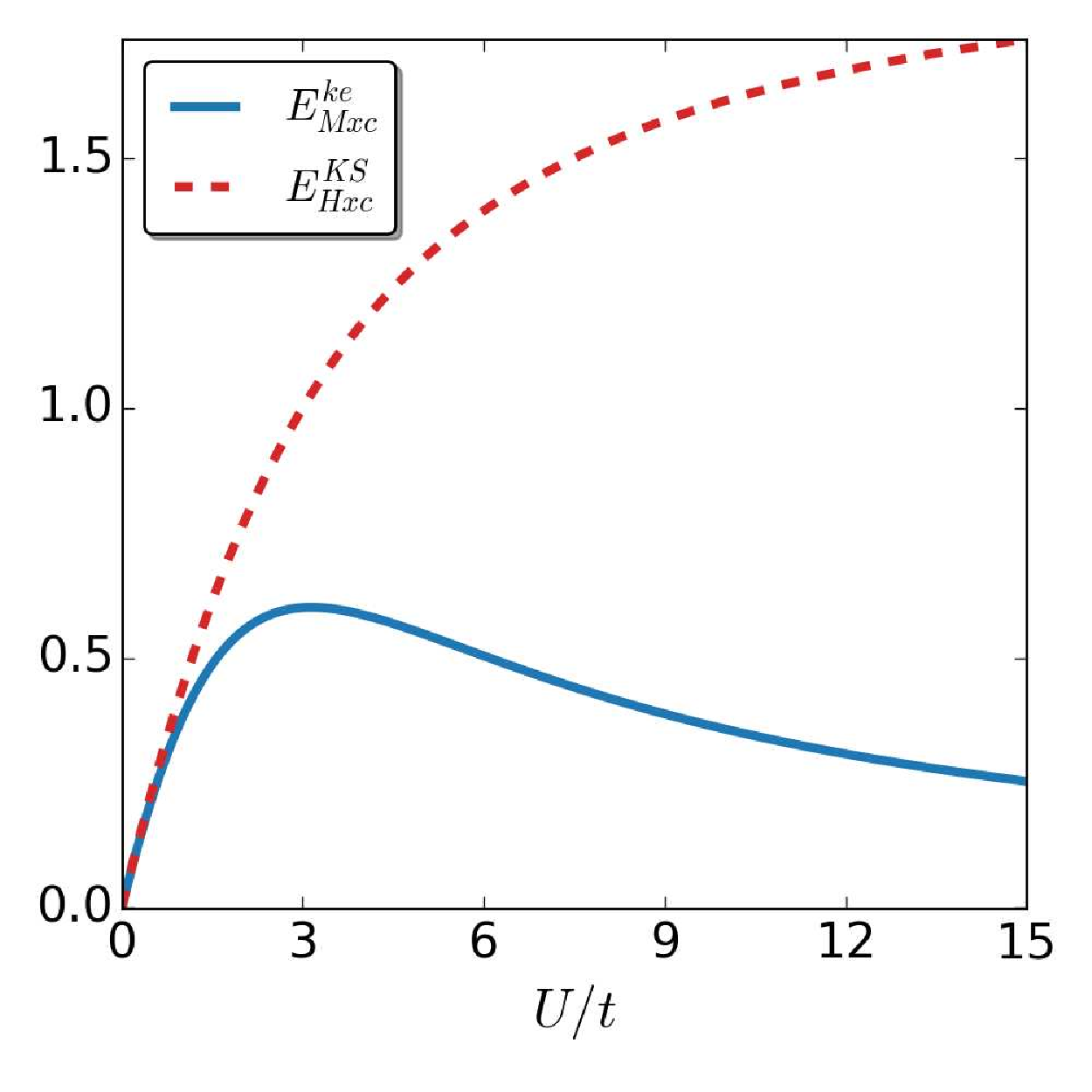}

\caption{Hxc energy $E_{Hxc}^{KS}$ (dashed line) and the corresponding energy term of the keKS system $E_{Mxc}^{ke}$ (continuous line) for a Hubbard dimer at half-filling with local potential $\Delta v/t=1$ as a function of $U/t$. We see that for $U>0$ it holds that $E_{Mxc}^{ke}<E_{Hxc}^{KS}$.}
\label{fig:EHxc}
\end{figure}

\begin{figure}
\includegraphics[width=0.95\columnwidth]{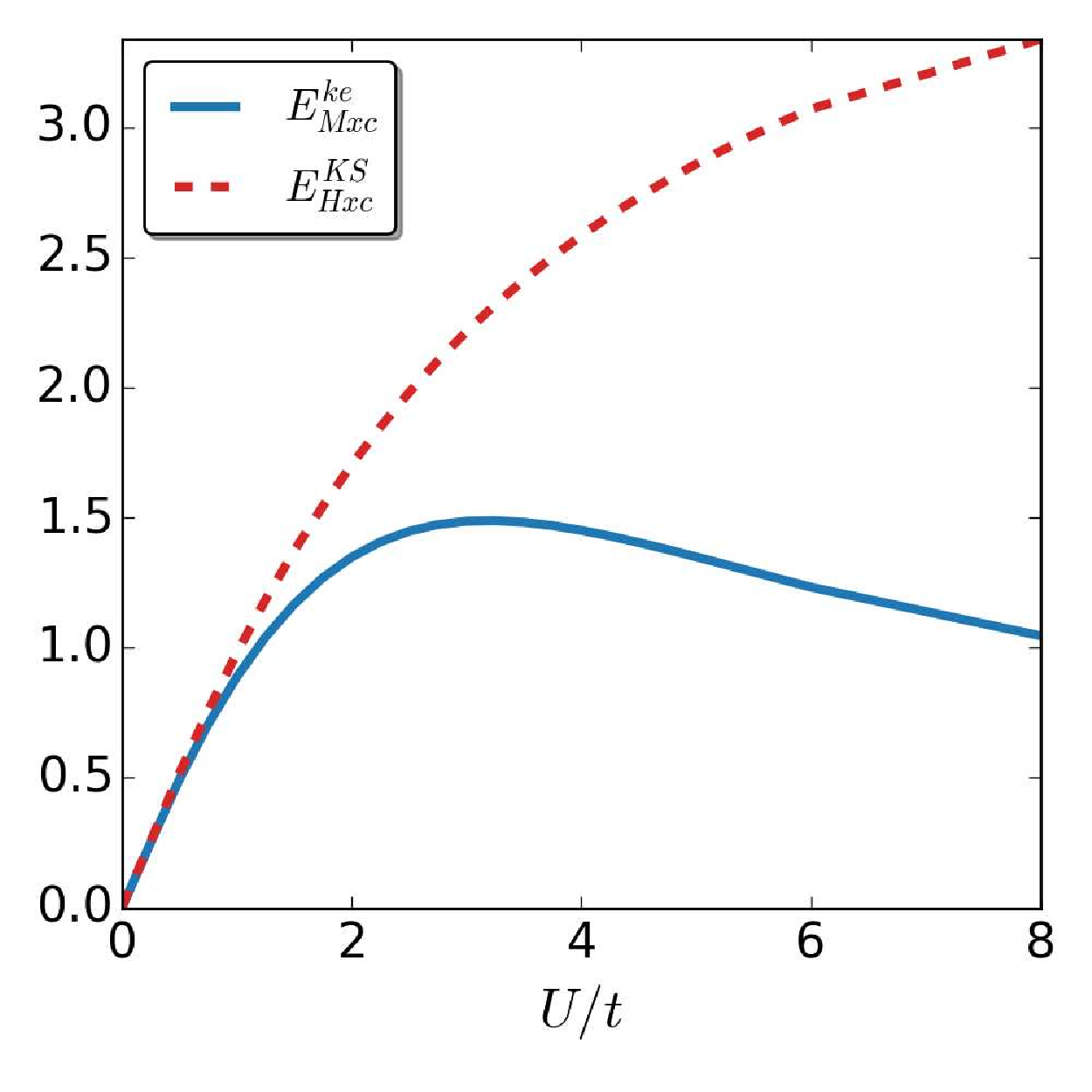}

\caption{Hxc energy $E_{Hxc}^{KS}$ (dashed line) and $E_{Mxc}^{ke}$ (continuous line) for a 4-site Hubbard model at half-filling with local potential $\Delta v_1/t=-\Delta v_3/t=0.625$ and $\Delta v_2/t=0.375$ as a function of $U/t$. We again find that for $U>0$ it holds that $E_{Mxc}^{ke}<E_{Hxc}^{KS}$.}
\label{fig:EHxc_4sites}
\end{figure}

Next we assess the practical implications of using the kinetic-energy density as basic functional variable along with the density. First, we use the construction of the exact keKS system and the corresponding KS system to compare the Hxc energy $E_{Hxc}^{KS}$ of the KS system with the corresponding quantity $E_{Mxc}^{ke}$ of the keKS system. This gives us a first indication of whether a keKS approach might help to capture also strong correlation effects more easily. For the KS system the Hxc energy is
\begin{eqnarray}
E_{Hxc}^{KS}=E_{gs}-\sum_{i=1}^{M}v_i n_i-\sum_{i=1}^{M-1}T_{i}^{KS},
\end{eqnarray}
where $T_{i}^{KS}=-2t\langle \Phi|\hat{\gamma}_{i,i+1}|\Phi\rangle$ is the kinetic energy density of the KS system and $|\Phi\rangle$ its ground state wavefunction. By $E_{gs}$ we denote the total ground-state energy of the interacting system and $v_i$ is its external potential. The corresponding energy contribution of the keKS system reads
\begin{eqnarray}
E_{Mxc}^{ke}=E_{gs}-\sum_{i=1}^{M}v_i n_i-\sum_{i=1}^{M-1}T_{i}^{ke}
\end{eqnarray}
where $T_{i}^{ke}=-2t^{ke}_i\langle \Phi^{ke}|\hat{\gamma}_{i,i+1}|\Phi^{ke}\rangle$.
Since the kinetic energy of the keKS system is identical to the interacting one by construction,
the $E_{Mxc}^{ke}\equiv E_{int}=U\sum_{i=1}^{M}\langle \Psi|\hat{n}^{\uparrow}_i \hat{n}^{\downarrow}_i|\Psi\rangle$ is equal only to the interaction energy in this case. The corresponding term of the KS system includes
kinetic-energy contributions as well. In Fig.~\ref{fig:EHxc} we plot $E_{Mxc}^{ke}$ and $E_{Hxc}^{KS}$ for a Hubbard dimer at half-filling with local potential $\Delta v/t=1$ as a function of the interaction strength $U/t$. Note that the data from the numerical inversion is used. Thus, both energy quantities are exact and there is no approximation involved. 
In Fig.~\ref{fig:EHxc_4sites}, we show the corresponding plot for 4-site Hubbard at half-filling with $\Delta v_1/t=-\Delta v_3/t=0.625$ and $\Delta v_2/t=0.375$ . These two systems for 2-site and 4-site will serve as our test systems and in the following we will refer to them as 2-site case and 4-site case, respectively.
As one can readily see in Figs.~\ref{fig:EHxc} and \ref{fig:EHxc_4sites} for every interaction strength $U> 0$ it holds that $E_{Mxc}^{ke}<E_{Hxc}^{KS}$. In the strong correlation limit, the kinetic-energy of the KS system is far from the interacting one. Having in mind the following relation
\begin{eqnarray}
E_{Mxc}^{ke}-E_{Hxc}^{KS}=T^{KS}-T^{ke},
\end{eqnarray}
it becomes apparent why $E^{ke}_{Mxc}$ and $E^{KS}_{Hxc}$ are so different for strong interactions. As a consequence the exchange-correlation potential derived from $E^{KS}_{Hxc}$ will need to take into account this difference in the strong interaction regime. 
In the keKS system, on the other hand, one needs to introduce a second field $\mathbf{t}^{Mxc}$ which is responsible for reproducing the kinetic-energy density along with the potential $\mathbf{v}^{Mxc}$ which ensures the density is reproduced.

\par

While the energy functionals are an interesting first indication that the keKS approach can be useful to treat also strongly-correlated systems, the real quantities of interest are the effective fields that the KS and keKS constructions employ. Especially those parts of the Hxc potential and of the Mxc hopping and potential that are not accessible by simple approximation strategies. Those parts, which one usually assumes to be small in practice, we will denote as correlation terms. Let us in the following, based on the EOMs we used to derive the iteration scheme, define parts of the effective fields that we can express explicitly in terms of the KS and keKS wave functions. Similar constructions based on the EOM of the density have been employed in DFT and TDDFT~\cite{local_force_Mx, fuks2016time}. For simplicity, we present the expressions only for the 2-site case. The expressions for 4-sites are given in App.~\ref{app:4sites_potentials}. The Hxc potential is defined as $\Delta v^{Hxc}[\mathbf{n}] =\Delta v^{s}[\mathbf{n}] - \Delta v[\mathbf{n}]$  (Eq.~\eqref{v_Hxc}), where $\mathbf{n}$ is the target density of the interacting system, $\Delta v^s$ is the local potential difference of the KS system, and $\Delta v$ is just the external potential of the interacting system. The EOMs for the non-interacting/interacting density,  \eqref{eq_mot_n_2sites}/\eqref{eq_mot_n_2sites_inter}, provide expressions for the local potential $\Delta v^{s}[\mathbf{n}]$ and $\Delta v[\mathbf{n}]$ of the non-interacting/interacting system.  Thus, the Hxc potential in the 2 site case reads
\begin{align}
\Delta v^{Hxc}[\mathbf{ n}]&=\frac{2t^2\Delta n}{T^{KS}}
-\frac{2t^2\Delta n}{T}-\frac{2Ut}{T}\sum_{\sigma\neq \sigma'}\langle \Psi|\hat{\gamma}_{1,2}^\sigma (\hat{n}_2^{\sigma'}-\hat{n}_1^{\sigma'})|\Psi\rangle.
\end{align}

We can decompose $\Delta v^{Hxc}$ in a Hartree-exchange part $\Delta v^{Hx}[\mathbf{n},\Phi]$  
\begin{align}
\Delta v^{Hx}[\mathbf{ n}, \Phi]=
-\frac{2Ut}{T^{KS}}
\sum_{\sigma \neq \sigma'}\langle \Phi |\hat{\gamma}_{1,2}^\sigma (\hat{n}_2^{\sigma'}-\hat{n}_1^{\sigma'})|\Phi\rangle,
\label{eq:exact-exchange}
\end{align}
which corresponds to the usual Hartree plus exchange approximation in standard DFT, and a remaining correlation part
\begin{align}
\Delta v_{c}^{KS}[\mathbf{ n},\Phi]&=
\frac{2t^2\Delta n}{T^{KS}}-
\frac{2t^2\Delta n}{T}\nonumber\\
&-\frac{2Ut}{T}\sum_{
\sigma \neq \sigma'}\langle \Psi |\hat{\gamma}_{1,2}^\sigma (\hat{n}_2^{\sigma'}-\hat{n}_1^{\sigma'})|\Psi\rangle\nonumber+\frac{2Ut}{T^{KS}}\sum_{
\sigma \neq \sigma'}\langle \Phi |\hat{\gamma}_{1,2}^\sigma(\hat{n}_2^{\sigma'}-\hat{n}_1^{\sigma'})|\Phi\rangle.
\label{eq:vc_ks_2sites}
\end{align}
Here we include the KS wave function in the functional dependencies to highlight that it is an orbital functional, i.e., it depends on the KS wave function. We note, however, that in the exact case the KS wave function is uniquely determined by the density. The above decomposition is similar to the one introduced in Ref.~\cite{local_force_Mx} and later used in, e.g., Refs.~\cite{fuks2016time,liao2017time}. In Eq.~\eqref{v_Mxc} we have
defined the Mxc potential $v^{Mxc}$ for the keKS system, which in the 2-site case (by using the same EOMs as before) reads
\begin{align}
\Delta v^{Mxc}[\mathbf{ n},T]=\frac{2{t^{ke}}^2 \Delta n}{T}
-\frac{2t^2\Delta n}{T}
-\frac{2Ut}{T}\sum_{\sigma \neq \sigma'}\langle \Psi |\hat{\gamma}_{1,2}^\sigma(\hat{n}_2^{\sigma'}-\hat{n}_1^{\sigma'})|\Psi\rangle.
\end{align}
We see that the first  two terms are completely determined by the keKS system, contrary to $\Delta v^{Hxc}$ of the KS system, where the second term cannot be given in terms of $\Delta n$ or $\Phi$ explicitly . One can identify a mean-field exchange potential, similarly to the Hartree-exchange potential of the KS system
\begin{align}
\Delta v^{Mx}[\mathbf{n},T,\Phi^{ke}]=\frac{2{t^{ke}}^2 \Delta n}{T}
-\frac{2t^2\Delta n}{T}-\frac{2Ut}{T}\sum_{
\sigma \neq \sigma'}\langle \Phi^{ke} | \hat{\gamma}_{1,2}^\sigma (\hat{n}_2^{\sigma'}-\hat{n}_1^{\sigma'})|\Phi^{ke}\rangle,
\label{eq:v_Mx}
\end{align}
which depends explicitely on the density, the kinetic-energy density and the ground-state of the keKS system. Let us at this point remark that if there is no approximation for the hopping parameter involved, i.e., when $t^{ke}=t$, the expression of $v^{Mx}$ in Eq.~\eqref{eq:v_Mx} is identical to the expression for $v^{Hx}$ in Eq.~\eqref{eq:exact-exchange}.  We note that for the exact case we consider here, i.e., at the solution point of the exact keKS non-linear equation, $t^{ke}$ can be explicitly given in terms of $t$ and the exact $\Phi^{ke}$. In practice, however, we do not know $t^{ke}[t, \textbf{n},T]=t + t^{Mxc}[\textbf{n},T]$ apriori and we need to include further an extra approximation for $t^{Mxc}[\textbf{n},T]$. Which approximations are possible (and how accurate they are) will be discussed next, and in the following section we will see how the practical form of $v^{Mx}[\mathbf{n},T,\Phi^{ke}]$, i.e., including an approximate $t^{Mxc}[\textbf{n},T]$, performs. The remaining local potential correlation term contains now only contributions from the difference in interaction
\begin{align}
\Delta v_{c}^{ke}[\textbf{n},T,\Phi^{ke}]=
-\frac{2Ut}{T}\sum_{
\sigma \neq \sigma'}\langle \Psi | \hat{\gamma}_{1,2}^\sigma (\hat{n}_2^{\sigma'}-\hat{n}_1^{\sigma'})|\Psi\rangle+\frac{2Ut}{T}\sum_{
\sigma \neq \sigma'}\langle \Phi^{ke} | \hat{\gamma}_{1,2}^\sigma (\hat{n}_2^{\sigma'}-\hat{n}_1^{\sigma'})|\Phi^{ke}\rangle.
\label{eq:vc_ke_2sites}
\end{align}
\begin{figure}
\includegraphics[width=0.95\columnwidth]{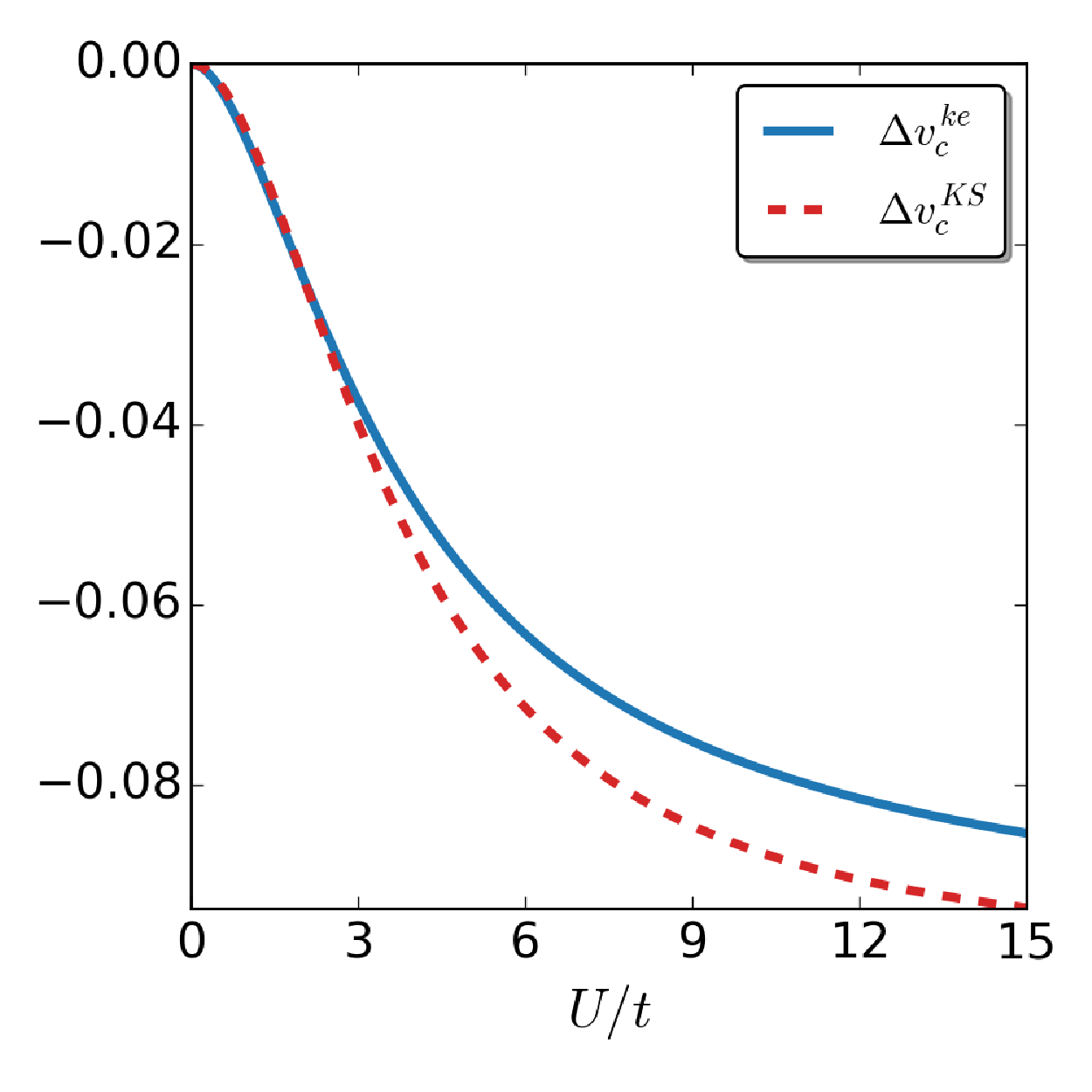}
\caption{The correlation potential of the keKS $\Delta v_c^{ke}$ (continuous curve) and KS system $\Delta v_c^{KS}$ (dashed curve) for the 2-site case as a function of the interaction strength U/t. Apart from a small region at vanishing interaction strength U, $|\Delta v_c^{ke}|<|\Delta v_c^{KS}|$.}
\label{fig:vc}
\end{figure}
\begin{figure}
\includegraphics[width=0.95\columnwidth]{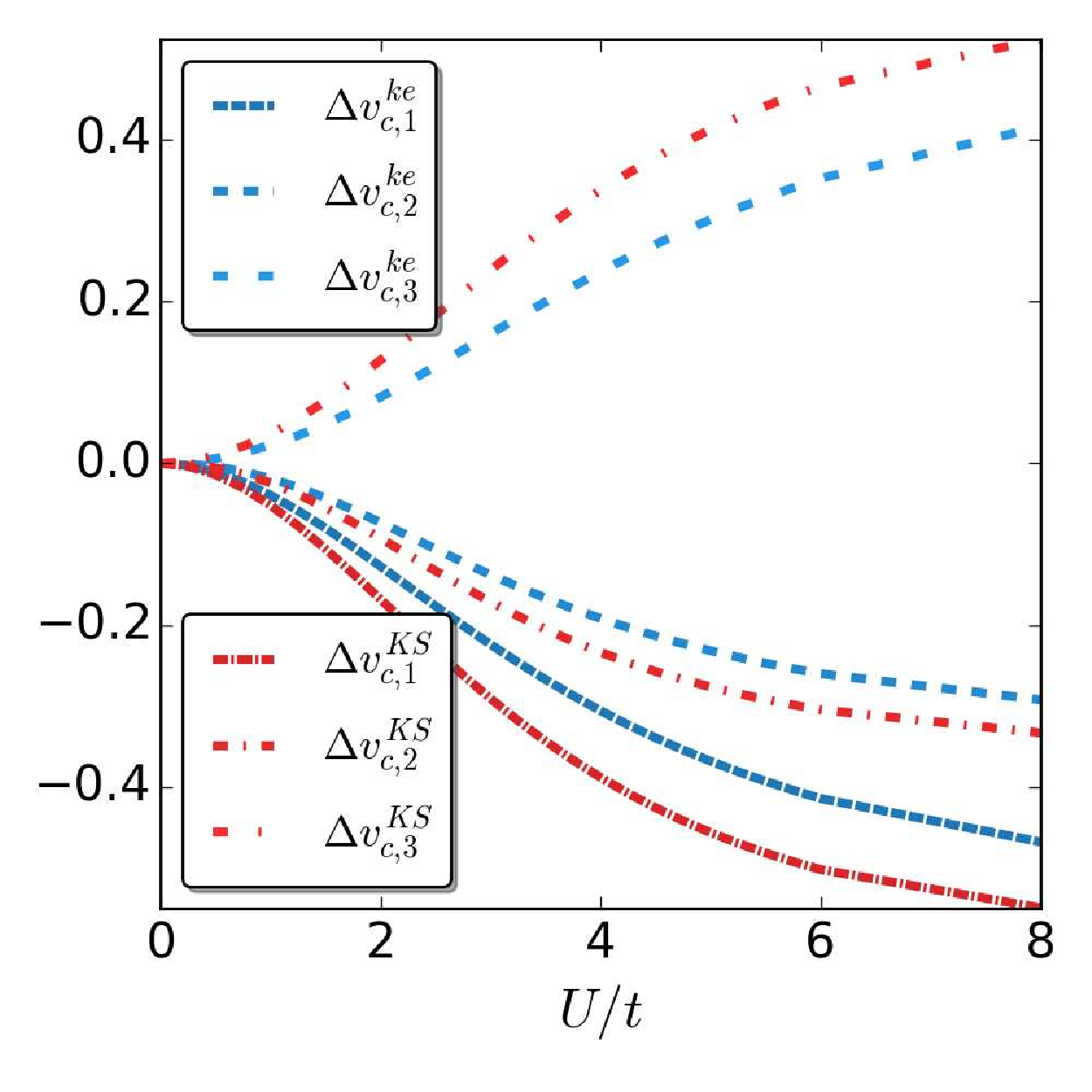}
\caption{The correlation potentials of the keKS $\Delta v_{c,i}^{ke}$ (dashed)  and KS system $\Delta v_{c,i}^{KS}$ (dotted-dashed) for the 4-site case as a function of the interaction strength U/t. Again we find that apart from a small region at vanishing interaction strength U, $|\Delta v_{c,i}^{ke}|<|\Delta v_{c,i}^{KS}|$.}
\label{fig:vc_4sites}
\end{figure}
In Fig.~\ref{fig:vc} we plot for the 2-site case the correlation KS and keKS potentials, which are given by Eqs.~\eqref{eq:vc_ks_2sites} and \eqref{eq:vc_ke_2sites} respectively. In Fig.~\ref{fig:vc_4sites} we plot the correlation potentials for the 4-site case, which are given for the KS system by Eqs.~\eqref{eq:vc_1^KS_4sites}-\eqref{eq:v^c_3_KS_4sites}
and for the keKS by
Eqs.~\eqref{eq:vc_1_ke_4sites}-\eqref{eq:vc_3_ke_4sites}.
As one can readily see for the 2-sites, the correlation potential is smaller in absolute value for the keKS system than in the KS one for all interaction strengths tested, apart from a small region at vanishing interaction. This follows from the fact that the kinetic contributions are included in the mean-field exchange potential $\Delta v^{Mx}_{i}$ in the keKS case. For the 4 sites we see the same trend. However, in the keKS construction we have a second effective field, which we so far did not take into account in our comparison. If we cannot find an analogous decomposition into an explicit and implicit part, where the explicit part can later be used as a functional approximation similar to $v^{Hx}_{i}$ in the KS case and $v^{Mx}_i$ as part of the keKS construction, then we did not gain anything. And if such an explicit part is introduced, we should check that it remains small compared to the unknown implicit part.

\par
\begin{figure}
\includegraphics[width=0.95\columnwidth]{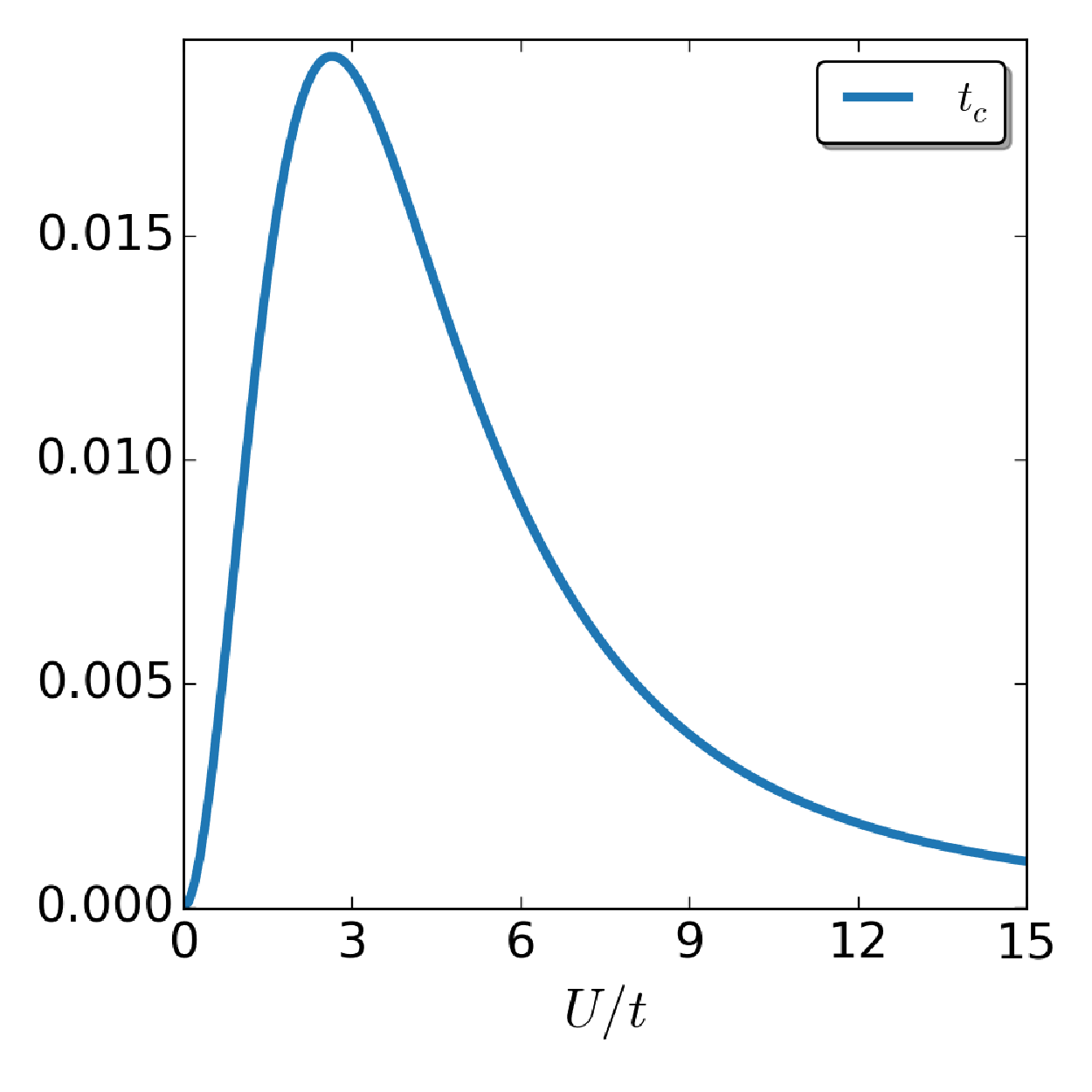}
\caption{Correlation part of the hopping $t_{c,i}$ in units of $t$, for the 2-site case as a function of interaction strength U/t. For strong interaction strength, the system resembles a homogeneous one so that the uniform type of approximation we employed becomes very good.}
\label{fig:2sites_tc}
\end{figure}
\begin{figure}
\includegraphics[width=0.95\columnwidth]{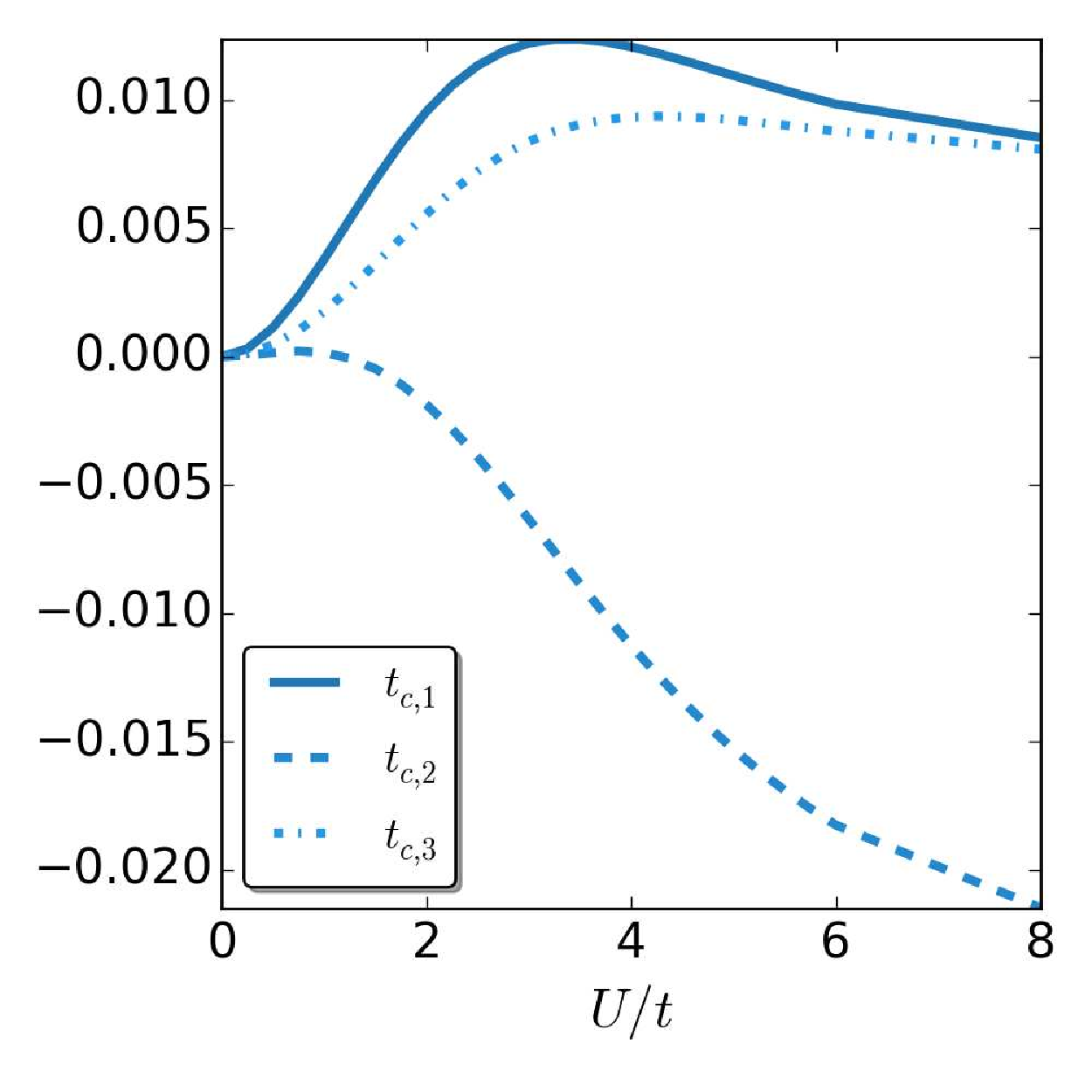}
\caption{Correlation part of the hopping $t_{c,i}$ in units of $t$, for the 4-site case as a function of interaction strength U/t. }
\label{fig:4site_tc}
\end{figure}

Starting from the fact that the kinetic-energy density $T_i$ of the interacting and the keKS system have to be the same by construction
\begin{eqnarray}
T_i=-2t^{ke}_i\gamma_{i,i+1}^{ke}=-2t\gamma_{i,i+1},
\label{eq:eq_Q}
\end{eqnarray}
subtracting from both sites of Eq.~\eqref{eq:eq_Q} $2t\gamma^{ke}_{i,i+1}$ and inserting the definition of $t^{Mxc}_i[\mathbf{ n},\mathbf{T}]\equiv t^{ke}_i[\mathbf{ n},\mathbf{T}]-t[\mathbf{n} ,\mathbf{T}]$, we get
that 
\begin{eqnarray}
t^{Mxc}_i[\mathbf{n},\mathbf{T}]=\frac{t[\mathbf{n},\mathbf{T}]\delta \gamma_i[\mathbf{n},\mathbf{T}]}{\gamma^{ke}_{i,i+1}[\mathbf{n},\mathbf{T}]},
\label{eq:tmxc}
\end{eqnarray}
where we have defined $\delta\gamma_{i}\equiv\gamma_{i,i+1}-\gamma^{ke}_{i,i+1}$. Up to here, there is no approximation involved. 
As the term $\delta \gamma_i$ involves the solution of an interacting and non-interacting problem, an approximation based on a reference solution suggest itself. The simplest such reference solution would be to use the homogeneous case of the interacting and the keKS system, respectively, similar to the local-density approximation in standard DFT. Since in the homogeneous case with periodic boundary conditions, as discussed in Sec.~\ref{sec:Mapping}, the keKS and the KS density matrices are the same, we can directly use well-known results such as the Bethe-ansatz solution at half filling. In this way it becomes also straightforward to extend the introduced approximation to the continuum case, where we can use reference calculations for interacting homogeneous continuum systems. Let us therefore define the explicit part of the hopping field by 
\begin{eqnarray}
t^{Mx}_i[\mathbf{n},\mathbf{T},\Phi^{ke}]=\frac{t\delta\gamma_i[n_i,t_i,U]}{\gamma^{ke}_{i,i+1}[\mathbf{n},\mathbf{T}]},
\label{eq:tmx}
\end{eqnarray}
where $\delta\gamma_i[n_i,t_i,U]=\gamma_{i,i+1}[n_i,t_i,U,\Delta \mathbf{v}=0]-\gamma^{KS}_{i,i+1}[n_i,t_i,\Delta \mathbf{v}=0]$. Here we assume that we have reference data for the homogeneous problems for different local hoppings $0<t_i$, local fillings $0<n_i<2$ and for the local interactions $0<U$. Further we ignore the dependence of the $t_i$ in the numerator on the internal pair $(\mathbf{n}, \mathbf{T})$ and use an explicit dependence on $\Phi^{ke}$ in the denominator. In the following we will simplify the explicit parts even further and will just take the homogeneous solution at half filling, i.e., $t^{Mx}_i[\mathbf{n},\mathbf{T},\Phi^{ke}] \equiv t^{Mx}_i[\mathbf{T},\Phi^{ke}]$, which we will take, however, from the respective 2-site and 4-site cases. In the 2-site case such an ansatz seems appropriate, since despite the zero-boundary conditions the keKS and the KS system are the same by construction. For the 4-site case, however, the zero-boundary conditions make the keKS and KS density matrices different. Hence the 4-site case is a very challenging test for the accuracy of such a simple approximation. In accordance to the above introduced approximation we will then define the correlation part of the local hopping field as $t_{c,i}[\mathbf{n},\mathbf{T},\Phi^{ke}]=t^{Mxc}_i[\mathbf{n},\mathbf{T}]-t^{Mx}_i[\mathbf{T},\Phi^{ke}]$.

In Fig.~\ref{fig:2sites_tc} we plot the correlation hopping field $t_c/t$ as a function of the interaction strength U/t for the 2-site case. We see that the value of $t_c$ is small compared to the chosen $t$ for all interaction strengths and especially for weak and strong interactions. For strong interactions the system resembles a homogeneous one as the interaction strength becomes more prominent in comparison to the local potential difference, thus $t_c$ becomes smaller and smaller in this regime. From this we can infer that for the case of a general system with periodic boundary conditions the homogeneous ansatz will capture not only the weak but also the strong-interaction limit accurately.
In Fig.~\ref{fig:4site_tc} we turn to the more challenging case of 4 sites with zero boundary conditions and plot the 3 different $t_{c,i}/t$ components as a function of the interaction strength. As one can readily see, all 3 $t_{c,i}$ are small for every interaction strength U. However, only two of them seem to converge to a value that is close to zero for strong interactions, at least for the parameter range we investigated\footnote{We want to point out, that for larger values of $U$ we encountered some convergence issues in the 4-site case. The reason being that while the density-matrices are not homogeneous, the density is, which causes some problems in the iteration scheme, where we divide by the density difference in each iteration. This problem, however, can be potentially overcome by using different update equations.}. Still, comparing the numerical values of the correlation hopping $t_{c,i}$ to the ones of the correlation potential $v_{c,i}^{ke}$, there is an order of magnitude difference. This gives some hope that crude approximations like the $t^{Mx}_{i}$ can still lead to accurate predictions. Let us test this for the 2-site case in the following section.

\begin{figure}
\includegraphics[width=0.95\columnwidth]{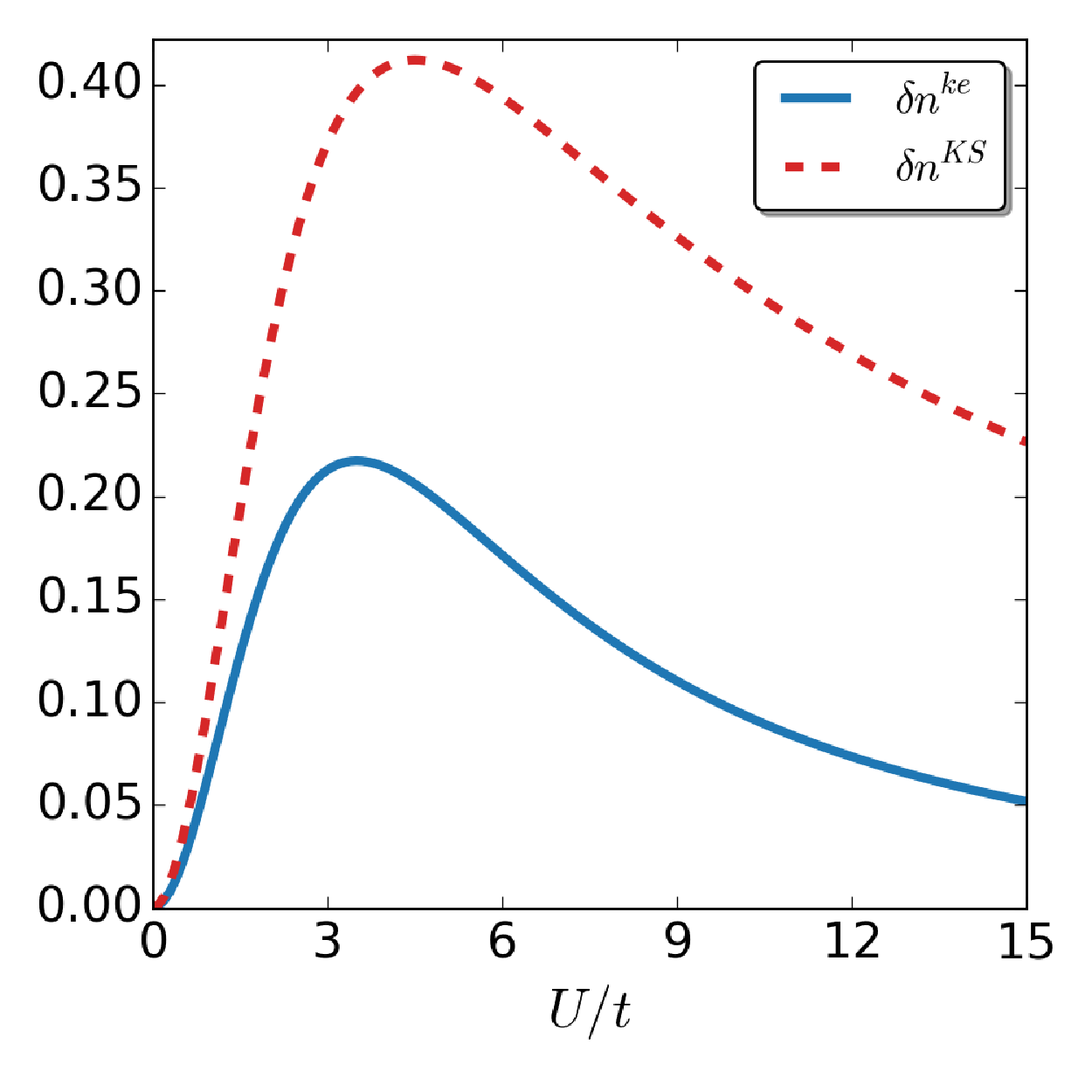}
\caption{The density difference $\delta n^{ke/KS}$ between the self-consistent calculations in the keKS system and the exact one (continuous, blue line), as well as for the self-consistent solution in the KS system and the exact one (dotted, red line), for the 2-site case as a function of interaction strength $U/t$.}
\label{fig:2sites_scf_n}
\end{figure}
\begin{figure}
\includegraphics[width=0.95\columnwidth]{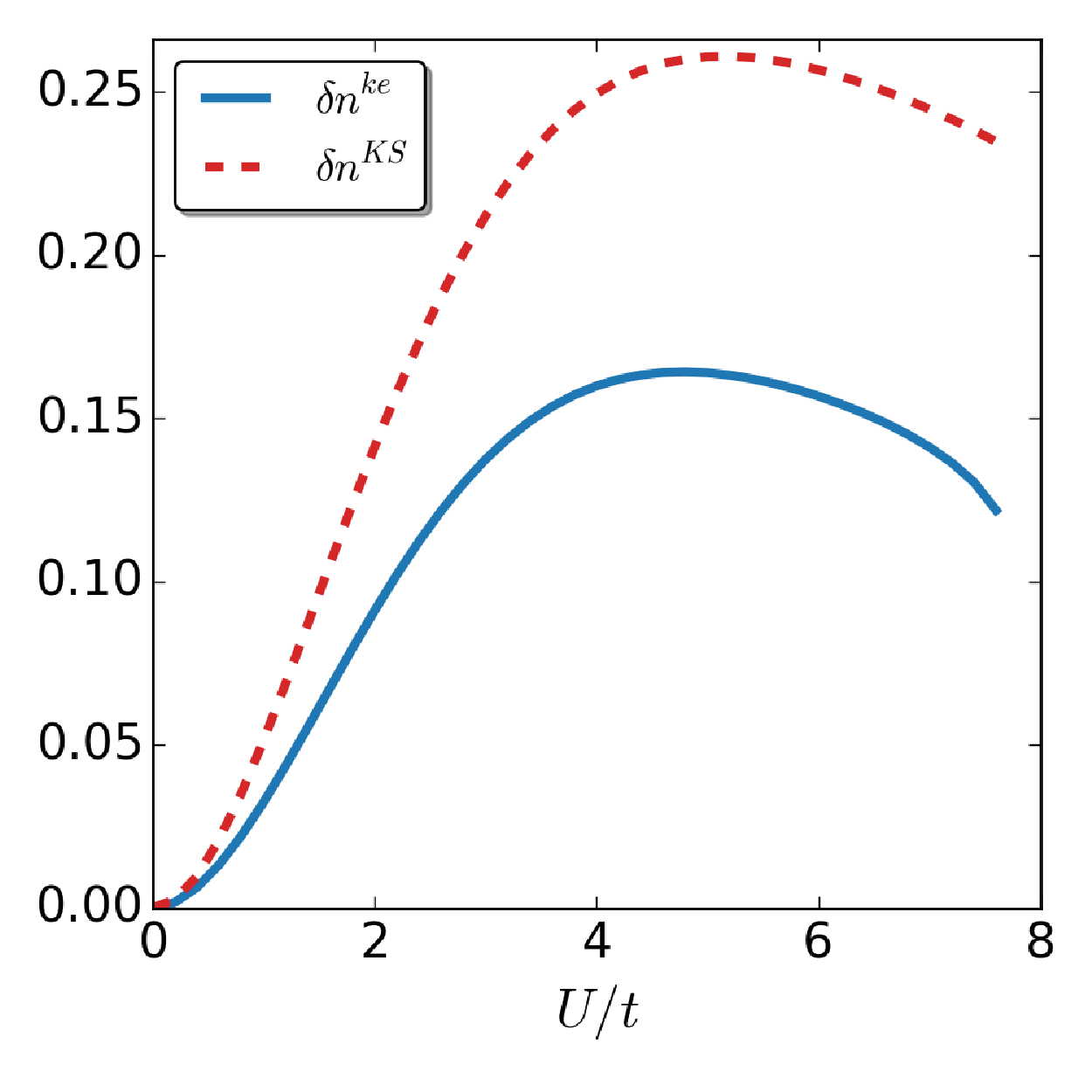}
\caption{The density difference $\delta n^{ke/KS}$ between the self-consistent calculations in the keKS system and the exact one (continuous, blue line), as well as for the self-consistent solution in the KS system and the exact one (dotted, red line), for the 4-site case as a function of interaction strength $U/t$.}
\label{fig:4sites_scf_n}
\end{figure}

\section{Comparing a self-consistent KS and keKS calculation}
\label{sec:SCF_Results}
\begin{figure}
\includegraphics[width=0.95\columnwidth]{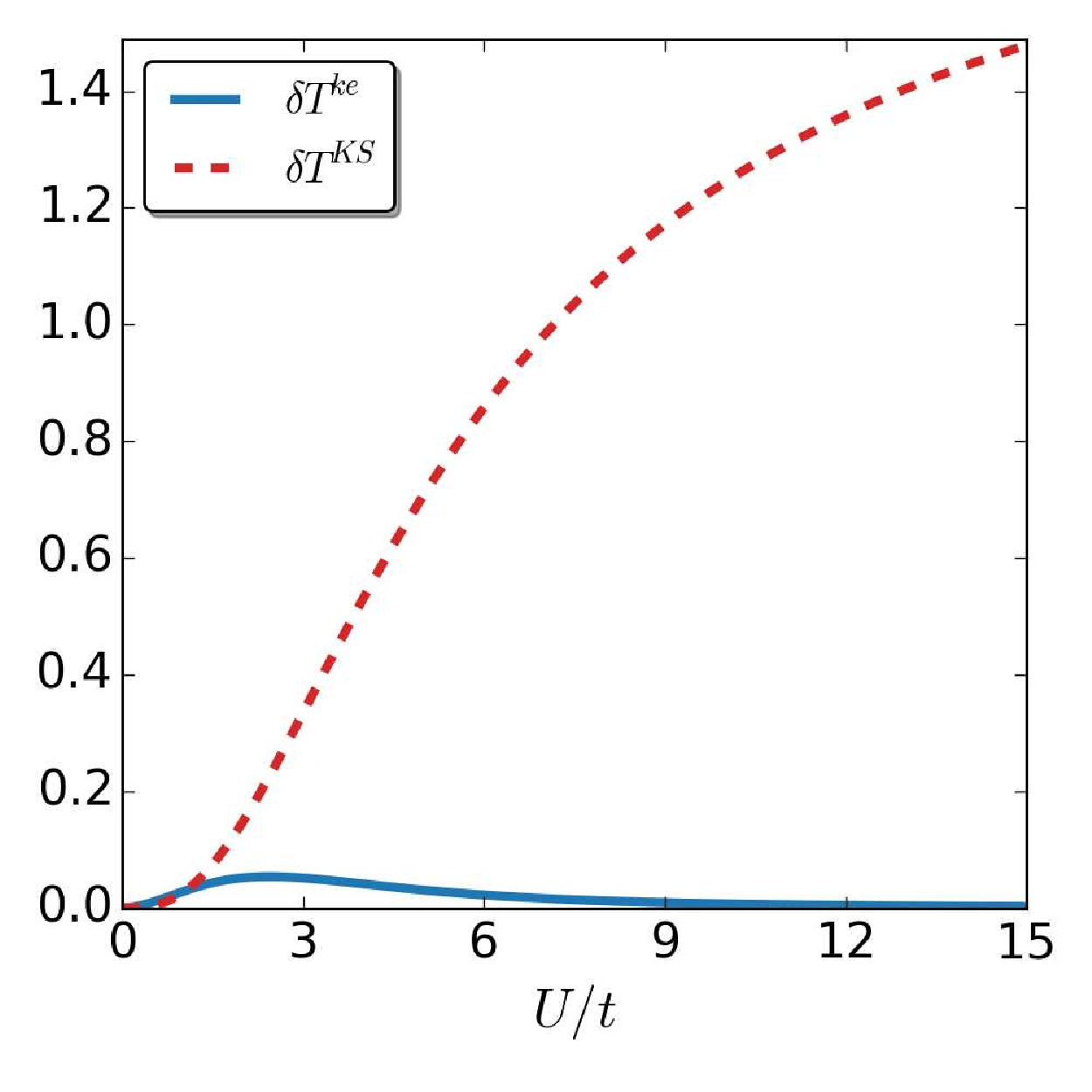}
\caption{The kinetic-energy density difference $\delta T^{ke/KS}$ between the self-consistent calculations in the keKS system and the exact one (continuous line), as well as for the self-consistent solution in the KS system and the exact one (dotted line), for the 2-site case as a function of interaction strength $U/t$.}
\label{fig:2sites_scf_T}
\end{figure}
\begin{figure}
\includegraphics[width=0.95\columnwidth]{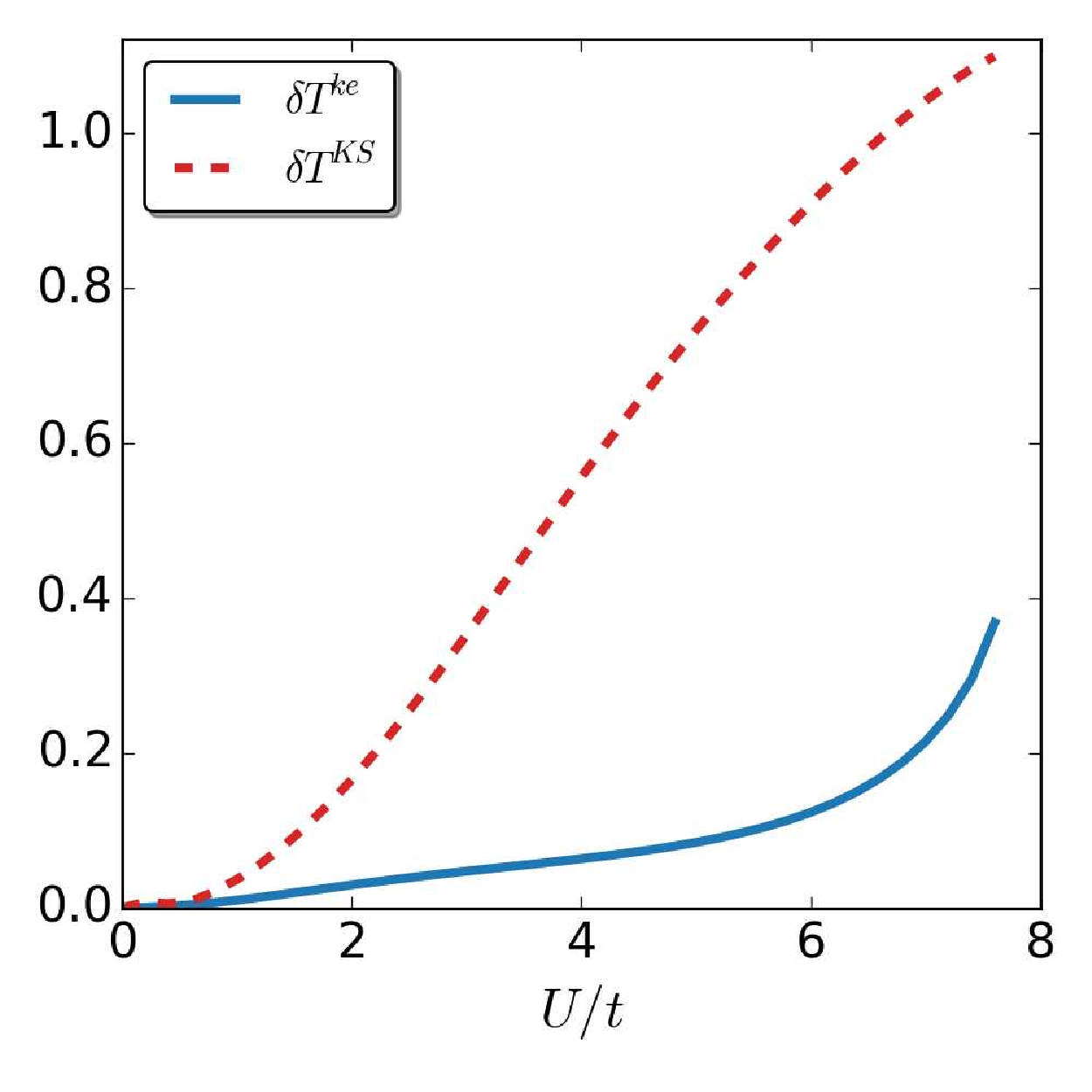}
\caption{The kinetic-energy density difference $\delta T^{ke/KS}$ between the self-consistent calculations in the keKS system and the exact one (continuous line), as well as for the self-consistent solution in the KS system and the exact one (dotted line), for the 4-site case as a function of interaction strength $U/t$.}. 
\label{fig:4sites_scf_T}
\end{figure}
Indeed, for both, the 2-site case (see Fig.~\ref{fig:2sites_scf_n}) as well as the more challenging 4-site case (see Fig.~\ref{fig:4sites_scf_n}), the self-consistent keKS approximation performs better than the corresponding self-consistent KS exact-exchange approximation. Since the main difference lies in the error correction to the local kinetic-energy density, we next also compare a measure for the difference in local kinetic-energy density: $\delta t^{ke/KS}=\sum_{i=1}^{M-1}|T^{ke/KS}_i-T_i|$, where $T^{ke/KS}_i$ is the kinetic-energy density between site $i$ and $i+1$ while $T_i$ is the corresponding interacting one. Not surprisingly, in both cases (see Figs.~\ref{fig:2sites_scf_T} and \ref{fig:4sites_scf_T}) the approximate kinetic energy density of the keKS system is much closer to the actual one than the bare KS energy density. While we see that for large interaction strengths the error is basically zero for the 2-site case, in the 4-site case again the issue with the boundary conditions becomes apparent. Nevertheless, for large systems (where issues at the boundaries will not be significant) or systems with periodic boundary conditions this issue will not arise and it can be expected that including the kinetic-energy density can help to treat multi-particle systems accurately from the weak to strong interaction regime.

While the above considerations about the exact correlation energies, potentials and hoppings are crucial to understand what the different approximations to the unknown exchange-correlation terms are able to capture, it is not their performance at the exact solutions that matters in practice. In practice we need to perform a self-consistent calculation with the approximate functionals, and it is not at all clear that such a calculation converges to a sensible solution or even converges at all. For instance, even for the prime example of a non-linear problem in quantum chemistry, i.e., the ground-state Hartree-Fock equation, the convergence to a unique solution has not be shown except for highly unusual cases~\cite{griesemer2012unique}. To finally test whether the proposed keDFT and its keKS construction can be used in practice to predict the properties of correlated many-electron systems, we perform self-consistent calculations for our 2-site and 4-site Hubbard models. We use the mean-field exchange approximation of Eq.~\eqref{eq:v_Mx} for 2-sites and of Eqs.~\eqref{eq:v^Mx_1_4sites}-\eqref{eq:v^Mx_3_4sites} for 4-sites together with the mean-field exchange approximation for the hopping term of Eq.~\eqref{eq:tmx} where we ignore the local density dependence in the numerator. This leads to 

\begin{eqnarray}
\left (\sum_{i=1}^{M-1} (t+t_i^{Mx}[\mathbf{T},\Phi^{ke}])(\hat{\gamma}_{i,i+1}+h.c.)
+\sum_{i=1}^M(v_i+ v_i^{Mx}[\mathbf{n},\mathbf{T},\Phi^{ke}])\hat{n} _i\right)
|\Phi^{ke}\rangle=\epsilon |\Phi^{ke}\rangle,
\label{eq:Scrh_aux_scf}
\end{eqnarray}
where we update the involved effective fields in every iteration until convergence is achieved. We then compare the densities and kinetic-energy densities that we get with the KS ones within the exact-exchange approximation (thus $t_i=t$ and $v^{Hx}$ given by Eq.~\eqref{eq:exact-exchange} for 2-sites and Eqs.~\eqref{eq:exact-exchange_1_4sites}-\eqref{eq:exact-exchange_3_4sites} for 4-sites). We do so as in this way we have for both the KS and the keKS construction the same level of approximation as $\mathbf{v}^{Mx}$ will reduce to $\mathbf{v}^{Hx}$ for $t_i=t$. This allows us to judge whether including the kinetic-energy density in the modeling of many-particle systems has any advantages over the usual density-only approach.

\par

We first quantify the density difference between the calculated quantities and the exact ones using the following measure: $\delta n^{ke/KS}=\sum_{i=1}^M|n^{ke/KS}_i-n_i|$, where
$n_i$ is the interacting density at site $i$ while $n^{ke/KS}_i$ is the corresponding density of the keKS/KS system.

\section{Conclusion and Outlook}
\label{sec:Outlook}

In this work, we have introduced a kinetic-energy density functional theory (keDFT) and the resulting kinetic-energy Kohn-Sham (keKS) scheme on a lattice. The idea was that by lifting the kinetic-energy density $\mathbf{T}$ to a fundamental variable along with the density $\mathbf{n}$, the resulting effective theory becomes easier to approximate since more parts are known explicitly. Since the new external field, a site-dependent hopping $\mathbf{t}$, is part of the kinetic-energy density, the usual Hohenberg-Kohn-type proof strategy to establish the necessary one-to-one correspondence between $(\mathbf{v}, \mathbf{t})$ and $(\mathbf{n}, \mathbf{T})$, where $\mathbf{v}$ is the usual on-site potential, does not work. However, besides giving proofs for specific cases and discussing the gauge freedom of the approach, we provided indication that the necessary bijectivity holds by numerically constructing the inverse maps from a given pair $(\mathbf{n}, \mathbf{T})$ to $(\mathbf{v}, \mathbf{t})$ for 2 to 4-site Hubbard models. We did so by introducing an iterative scheme based on the equations of motion (EOMs) of the density and the kinetic-energy density. Based on these EOMs, we then introduced a decomposition of the two unknown effective fields of the keKS scheme, the mean-field-exchange-correlation potential $v_{i}^{Mxc}[\mathbf{n}, \mathbf{T}]$ and the mean-field exchange-correlation hopping $t^{Mxc}[\mathbf{n}, \mathbf{T}]$, into explicitly known mean-field exchange and unknown correlation parts. By comparing the unknown parts of the standard Kohn-Sham (KS) approach to the keKS approach we saw that including the kinetic-energy density in the fundamental variables reduced the unknown parts considerably. Finally, we tested the keKS approach in practice by solving the resulting non-linear equations with the introduced mean-field exchange approximations. We found that the mean-field exchange keKS outperforms the corresponding exact-exchange KS from weak to strong interactions and hence holds promise to become an alternative approach to treat many-particle systems efficiently and accurately.\par

While the presented approach was thoroughly investigated only for simple few-sites problems, its extension to many sides, arbitrary dimensions and even the continuum is straightforward. The main reason why the keKS scheme can be more accurate than the usual KS scheme is that we model explicitly the kinetic-energy density. Since the simple kinetic-energy density approximations we introduced proved to be already quite reasonable, the extension to the continuum seems especially promising. For homogeneous systems many reference calculations exist that can be used to derive a universal local kinetic-energy density approximation. But following the presented route, we will in a follow-up work first consider the accuracy of the keKS scheme based on the mean-field exchange approach for many sites in one and two dimensions before going to the continuum. It is still easier to calculate and to compare to exact solutions in the lattice case. 

\section*{Acknowledgments}

Financial support from the European Research Council (ERC-2015-AdG-694097), by the European Union’s H2020 program under GA no.676580 (NOMAD) is acknowledged. F. G. E. has received funding from the European Union’s Framework Programme for Research and Innovation Horizon 2020 (2014-2020) under the Marie Sk{\l}odowska-Curie Grant Agreement No. 701796.

\begin{appendix}
\section{Two sites proof of $( \Delta v, t) \leftrightarrow
(\Delta n, T)$}

\label{app:proof_2_sites}
In this Appendix we provide a proof of the bijectiveness of the mapping between density and kinetic energy density and the corresponding fields in the case of a non-interacting system of up to $N\leq 3$ electrons on a two-site lattice.
For this case, the non interacting Hamiltonian reads 
\begin{eqnarray}
\hat{H}=-t(\hat{\gamma}_{1,2}+h.c.)
+\frac{\Delta v}{2} (\hat{n_1}-\hat{n_2})
\end{eqnarray}
where $\Delta v \in \mathbb{R}$
and $t > 0$.
The mapping that we wish to show that is bijective is the following one
\begin{eqnarray}
(t, \Delta v) \overset{1:1}{\leftrightarrow}
(T, \Delta n)
\label{map2sites}
\end{eqnarray}
with $ T > 0$ and $\Delta n \in ]-N, N[$. Here $\Delta v\equiv v_1-v_2$ and $\Delta n \equiv n_1-n_2$, where the lower indexes refer to different site points.
Note that $t \geq 0$ is a certain gauge choice as 
$\{(t, \Delta v), (-t,\Delta v)\}\mapsto (T, \Delta n)$. Moreover, due to particle hole symmetry it holds also that $\{(-t,-\Delta v), (t,-\Delta v)\}\mapsto (T, \Delta n)$. Thus from now on, when we refer to different potentials and hopping parameters,  we will mean that they differ by more than a sign change. For the local potential the gauge choice is $v_1+v_2=0$ as throughout the document.

\vspace*{3\baselineskip}

Proof:
We are going to prove \eqref{map2sites} through different cases.

\underline{Case 1}: Two Hamiltonians $\hat{H}$, $\hat{H'}$ have the same hopping parameters
$t=t'$ but different local potentials $\Delta v \neq \Delta v'$. 

Then from the Hohenberg-Kohn theorem, we have that the corresponding wavefunctions are different, i.e. $\Psi \neq \Psi'$ and the same holds also for the densities $\Delta n \neq \Delta n'$ and consequently $(T, \Delta n)\neq (T', \Delta n')$.

\underline{Case 2}: Two Hamiltonians $\hat{H}$, $\hat{H'}$ have different hopping parameters but the same local potentials 
$t \neq t'$ but $\Delta v \neq \Delta v'$.

Assume that $(T,\Delta n)= (T, \Delta n')$. Then we have two wavefunctions $\Psi$ and $\Psi'$ that are ground states of the corresponding Hamiltonians 
\begin{align}
\hat{H}\Psi &=E \Psi\nonumber\\
\hat{H'}\Psi'&=E' \Psi'
\end{align}
We can multiply $\hat{H'}$ with a scaling factor $\lambda=\frac{t}{t'}$ so that we get a Hamiltonian $\hat{H''}$ which has the same hopping as the $\hat{H}$.

\begin{eqnarray}
\hat{H''}\Psi'=\lambda \hat{H'}\Psi =\lambda E' \Psi' \Rightarrow\nonumber\\
\left( -t\sum_{\sigma=\uparrow,\downarrow}(\hat{c}_{1}^{\sigma\dagger}\hat{c}_{2}^{\sigma}+h.c.)
+\frac{t}{t'} \frac{\Delta v}{2} (\hat{n_1}-\hat{n_2})\right)\Psi'=\lambda E'\Psi'
\end{eqnarray}

By assumption $\Delta n = \Delta n'$, which means that we have two different ground state wavefunctions with different local potentials $\Delta v \neq \frac{t}{t'} \Delta v$ which still give the same densities. This is clearly in contradiction with Case 1.

\underline{Case 3}: Two Hamiltonians $\hat{H}$, $\hat{H'}$ have different hopping parameters  
$t \neq t'$ and different local potentials $\Delta v \neq \Delta v'$. 

Again, we assume that $(T, \Delta n)= (T', \Delta n')$ and we scale such that we find:

\begin{eqnarray}
\left( -t \sum_{\sigma=\uparrow,\downarrow}(\hat{c}_{1}^{\sigma\dagger}\hat{c}_{2}^{\sigma}+h.c.)
+\frac{t}{t'} \frac{\Delta v'}{2} (\hat{n_1}-\hat{n_2})\right)\Psi'=\lambda E'\Psi'
\end{eqnarray}

Requiring that $\Delta n=\Delta n' $ can only hold if also the local potentials are the same as we showed in Case 1. Thus, it has to hold that
\begin{eqnarray}
\Delta v=\frac{t}{t'}\Delta v'=\lambda \Delta v'
\end{eqnarray}

which means that the two Hamiltonians $\hat{H}$ and $\hat{H'}$ are connected through the scaling relation
$\hat{H}=\lambda \hat{H'}$.
Thus we have
\begin{eqnarray}
\hat{H}=\lambda \hat{H'}\Rightarrow \Psi=\Psi'\Rightarrow T \neq T' 
\end{eqnarray}

which contradicts our initial assumption.

\section{Equations of Motion}
\label{app:eqs_motion}
In this Appendix we derive the EOMs that we use in our numerical inversion scheme and to define the $\mathbf{v}^{Hxc}$ and $\mathbf{v}^{Mxc}$ potentials. Furthermore, we provide the explicit expressions of $\mathbf{v}^{Hxc}$ and $\mathbf{v}^{Mxc}$ for 4 sites. Finally, we discuss how the number of useful EOMs for our numerical inversion scheme depends on the number of sites.\par
The EOM for a generic operator $A$, which has no explicit time dependence, is given by
\begin{eqnarray}
\dot{\hat{A}}=i[\hat{H}, {\hat{A}}]~.
\end{eqnarray}
We are interested in obtaining EOM for a non interacting Hamiltonian of type \eqref{ke_Ham}. For a state $\Psi$ it follows that
\begin{equation}
\dot{A}= i \langle \Psi|[\hat{H}, {\hat{A}}]|\Psi\rangle ~,
\label{eq_mot_Ai}
\end{equation}
which is an EOM for the observable associated to the operator $\hat{A}$. When $\Psi$ is the ground state \eqref{eq_mot_Ai} equals zero.
Let's take now this operator to be the density. Since the first order EOM for the density, i.e. the continuity equation~\eqref{continuity}, is trivially satisfied as the current is just zero in the ground state, we are consider the second time derivative of the density:
\begin{eqnarray}
\ddot{n}_i=-\mathcal{D}_{-}\left(\mathcal{D}\Upsilon_i-2t^2_i \mathcal{D}n_i+
(\mathcal{D} v_i)T_i
\right),
\label{eq_mot_ni}
\end{eqnarray}
where we have introduced the forward  (backward) difference operators $\mathcal{D}$ ($\mathcal{D}_-$) which act on one index objects:$\mathcal{D}f_i=f_{i+1}-f_{i}$ , $\mathcal{D}_{-}f_i=f_i-f_{i-1}$. 
In Eq.~\eqref{eq_mot_ni} we furthermore introduced
\begin{eqnarray}
\Upsilon_i=t_it_{i+1}(\gamma_{i+1,i-1}+c.c.),
\end{eqnarray}
which, by analogy to the continuum case, can be identified as the kinetic contribution to the momentum-stress tensor. The time derivative of the kinetic energy density also leads to
\begin{eqnarray}
\dot{T_i}=-\mathcal{D} \Xi_i+(\mathcal{D}v_i)J_i,
\label{eq_mot_Ti_1}
\end{eqnarray}
where we introduced the kinetic-energy current
\begin{eqnarray}
\Xi_i=it_it_{i-1}(\gamma_{i+1,i-1}-c.c.) ~. \label{xi_def}
\end{eqnarray}
Both $\Xi_i$ and $J_i$ vanish trivially for real-valued wave functions such as the ground state, so Eq.\ \eqref{eq_mot_Ti_1} is fulfilled trivially. Taking yet another time derivative leads to
\begin{eqnarray}
\ddot{T_i}=-\mathcal{D} \dot{\Xi}_i+(\mathcal{D}v_i)\dot{J}_i ~.
\label{eq_mot_Ti_2}
\end{eqnarray}
In~\eqref{eq_mot_Ti_2}, there are
two more EOMs involved. The one for $\Xi_i$ is
\begin{eqnarray}
\dot{\Xi}_i=-\mathcal{D}_{-}\Lambda_i-\left((\mathcal{D} +
\mathcal{D}_{-})v_i\right)\Upsilon_i
-t_i^2\mathcal{D}_{-}T_i
+(\mathcal{D}_{-}t_i^2)T_i,
\end{eqnarray}
with
\begin{eqnarray}
\Lambda_i=t_{i+1}t_it_{i-1}(\gamma_{i+2,i-1}+c.c.).
\end{eqnarray}
And the EOM for the current is
\begin{eqnarray}
\dot{J}_i=\mathcal{D}\Upsilon_i-2t_i^2\mathcal{D}n_i+
(\mathcal{D} v_i)T_i.
\end{eqnarray}
Apart from the EOM that are derived from a non-interacting Hamiltonian we use the second order EOM for the interacting density to define the  $\mathbf{v}^{Hxc}$ and $\mathbf{v}^{Mxc}$ potentials

\begin{eqnarray}
\ddot{n}_i=-\mathcal{D}_{-}\left(\mathcal{D}\Upsilon_i-2t^2_i \mathcal{D}n_i+
(\mathcal{D} v_i)T_i
\right)
+2Ut\sum_{i=1}^{M-1}\sum_{\sigma \neq \sigma'}\langle \Psi |\hat{\gamma}_{i,i+1}^{\sigma} (\hat{n}_{i+1}^{\sigma'}-\hat{n}_i^{\sigma'})|\Psi\rangle ~.
\label{eq_mot_ni_inter}
\end{eqnarray}
 As an example, let us show here the EOM for $\ddot{n}_1$ for the two site interacting Hamiltonian, 
\begin{align}
\label{eq_mot_n_2sites_inter}
\ddot{n}_1= 
2 t^2\Delta n_1
-\Delta v_1 T_1 
+2Ut\sum_{
\sigma \neq \sigma'}\langle \Psi |\hat{c}_1^{\sigma\dagger}\hat{c}_2^{\sigma} (\hat{n}_2^{\sigma'}-\hat{n}_1^{\sigma'})|\Psi\rangle. 
\end{align}
\par

In order to find the effective fields for the keKS system, we solve Eq.~\eqref{eq_mot_ni} and Eq.~\eqref{eq_mot_Ti_2} iteratively for $v_i$ and $t_i$ using the exact results for the density and kinetic-energy density. Formally, Eq.~\eqref{eq_mot_ni} defines $M$ equations, i.e. one for every site. However, since the sum of densities at every site has to give the total number of electrons, it holds that:
\begin{eqnarray}
\sum_{i=1}^{M}\ddot{n}_i=0
\label{sum_ni}
\end{eqnarray}
i.e. we get $M-1$ non trivial equations.
Now for M sites we do have $M-1$ different $T_i$ of the form Eq.~\eqref{eq_mot_Ti_2}.
But as the total energy of the $keKS$ system is fixed to some value $E$, i.e.
\begin{eqnarray}
\sum_{i=1}^{M-1}T_i+\sum_{i=1}^M v_i n_i=E
\label{eq:energy_nonint}
\end{eqnarray}
taking the second time derivative of the above equation shows us that we have $M-2$ non trivial equations for $T_i$. Taking into account the gauge choice of $v_i$ (Eq.~\eqref{gauge_v}) and the particle number conservation Eq.~\eqref{sum_ni} we see that the equations for $\ddot n_i$ and $\ddot T_i$ connect through the following relation
\begin{eqnarray}
\sum_{i=1}^{M-1}\ddot{T}_i+\sum_{i=1}^{M-1} v_i \ddot{n}_i
+\sum_{i=1}^{M-1}v_i\sum_{j=1}^{M-1}\ddot{n}_j
=0.
\label{n_i_Q_i_relation}
\end{eqnarray}
\par
At this point we have to mention that Eq.~\eqref{sum_ni} still holds for the interacting system, however
Eqs.~\eqref{eq:energy_nonint},~\eqref{n_i_Q_i_relation} will not hold anymore since the on site repulsion term enters the energy expression in this case.

\section{Exchange-correlation potentials for four sites} \label{app:4sites_potentials}

From the EOMs \eqref{eq_mot_ni} and\eqref{eq_mot_ni_inter} one can derive the $\mathbf{v}^{Hxc}$ and $\mathbf{v}^{Mxc}$ for any number of sites. We give here their expressions for 4-sites since it is one of our test cases in the current manuscript.
\begin{eqnarray}
\Delta v_1^{Hxc}
[\mathbf{ n}]=\frac{2t^2\Delta n_1+
2t^2\gamma^{KS}_{13}}{T_1^{KS}}
+\frac{-2t^2\Delta n_1-2t^2\gamma_{13}}{T_1}
-\frac{2Ut}{T_1}\sum_{\sigma\neq \sigma'}\langle \Psi|\hat{\gamma}_{1,2}^\sigma (\hat{n}_2^{\sigma'}-\hat{n}_1^{\sigma'})|\Psi\rangle
\label{eq:v^Hxc_1_4sites}
\end{eqnarray}
\begin{align}
\Delta v_2^{Hxc}
[\mathbf{ n}]&=\frac{2t^2\Delta n_2-2t^2\gamma^{KS}_{13}+2t^2\gamma^{KS}_{24}}{T_2^{KS}}
+\frac{-2t^2\Delta n_2
+2t^2\gamma_{13}
-t^2\gamma_{24}}{T_2}\nonumber\\
&-\frac{2Ut}{T_2}\sum_{\sigma\neq \sigma'}\langle \Psi|\hat{\gamma}_{2,3}^\sigma (\hat{n}_3^{\sigma'}-\hat{n}_3^{\sigma'})|\Psi\rangle
\label{eq:v^Hxc_2_4sites}
\end{align}
\begin{eqnarray}
\Delta v_3^{Hxc}
[\mathbf{ n}]=\frac{2t^2\Delta n_3
-2t^2\gamma^{KS}_{24}}{T_3^{KS}}+
\frac{-2t^2\Delta n_3+2t^2\gamma_{24}}{T_3}
-\frac{2Ut}{T_3}\sum_{\sigma\neq \sigma'}\langle \Psi|\hat{\gamma}_{3,4}^\sigma (\hat{n}_4^{\sigma'}-\hat{n}_3^{\sigma'})|\Psi\rangle.
\label{eq:v^Hxc_3_4sites}
\end{eqnarray}
Similarly to the 2-site case we can decompose $\mathbf{v}^{Hxc}$ to a $\mathbf{v}^{Hx}$ and a $\mathbf{v}^{c}$ contribution:
\begin{align}
\Delta v^{Hx}_1[\mathbf{ n}, \Phi]=
-\frac{2Ut}{T^{KS}_1}
\sum_{\sigma \neq \sigma'}\langle \Phi |\hat{\gamma}_{1,2}^\sigma (\hat{n}_2^{\sigma'}-\hat{n}_1^{\sigma'})|\Phi\rangle
\label{eq:exact-exchange_1_4sites}
\end{align}
\begin{align}
\Delta v^{Hx}_2[\mathbf{ n}, \Phi]=
-\frac{2Ut}{T^{KS}_2}
\sum_{\sigma \neq \sigma'}\langle \Phi |\hat{\gamma}_{2,3}^\sigma (\hat{n}_3^{\sigma'}-\hat{n}_2^{\sigma'})|\Phi\rangle
\label{eq:exact-exchange_2_4sites}
\end{align}
\begin{align}
\Delta v^{Hx}_3[\mathbf{ n}, \Phi]=
-\frac{2Ut}{T^{KS}_3}
\sum_{\sigma \neq \sigma'}\langle \Phi |\hat{\gamma}_{3,4}^\sigma (\hat{n}_4^{\sigma'}-\hat{n}_3^{\sigma'})|\Phi\rangle
\label{eq:exact-exchange_3_4sites}
\end{align}
\begin{align}
\Delta v_{c,1}^{KS}[\mathbf{ n},\Phi]&=
\frac{2t^2\Delta n_1
+
2t^2\gamma^{KS}_{13}}{T_1^{KS}}
+\frac{-2t^2\Delta n_1
-2t^2\gamma_{13}}{T_1}
\nonumber\\
&-\frac{2Ut}{T_1}\sum_{
\sigma \neq \sigma'}\langle \Psi |\hat{\gamma}_{1,2}^\sigma (\hat{n}_2^{\sigma'}-\hat{n}_1^{\sigma'})|\Psi\rangle+\frac{2Ut}{T^{KS}_1}\sum_{
\sigma \neq \sigma'}\langle \Phi |\hat{\gamma}_{1,2}^\sigma(\hat{n}_2^{\sigma'}-\hat{n}_1^{\sigma'})|\Phi\rangle
\label{eq:vc_1^KS_4sites}
\end{align}
\begin{align}
\Delta v_{c,2}^{KS}
[\mathbf{ n},\Phi]&=\frac{2t^2\Delta n_2-2t^2\gamma^{KS}_{13}+2t^2\gamma^{KS}_{24}}{T_2^{KS}}
+\frac{-2t^2\Delta n_2
+2t^2\gamma_{13}
-t^2\gamma_{24}}{T_2}\nonumber\\
&-\frac{2Ut}{T_2}\sum_{\sigma\neq \sigma'}\langle \Psi|\hat{\gamma}_{2,3}^\sigma (\hat{n}_3^{\sigma'}-\hat{n}_3^{\sigma'})|\Psi\rangle
+
\frac{2Ut}{T_2^{KS}}\sum_{\sigma\neq \sigma'}\langle \Phi|\hat{\gamma}_{2,3}^\sigma (\hat{n}_3^{\sigma'}-\hat{n}_3^{\sigma'})|\Phi\rangle
\label{eq:v^c_2_KS_4sites}
\end{align}

\begin{align}
\Delta v_{c,3}^{KS}
[\mathbf{ n},\Phi]&=\frac{2t^2\Delta n_3
-2t^2\gamma^{KS}_{24}}{T_3^{KS}}+
\frac{-2t^2\Delta n_3+2t^2\gamma_{24}}{T_3}\nonumber\\
&-\frac{2Ut}{T_3}\sum_{\sigma\neq \sigma'}\langle \Psi|\hat{\gamma}_{3,4}^\sigma (\hat{n}_4^{\sigma'}-\hat{n}_3^{\sigma'})|\Psi\rangle
+\frac{2Ut}{T_3^{KS}}\sum_{\sigma\neq \sigma'}\langle \Phi|\hat{\gamma}_{3,4}^\sigma (\hat{n}_4^{\sigma'}-\hat{n}_3^{\sigma'})|\Phi\rangle.
\label{eq:v^c_3_KS_4sites}
\end{align}

Next we give the corresponding expressions for the keKS system:

\begin{eqnarray}
\Delta v_1^{Mxc}
[\mathbf{ n}]=\frac{2t_1^2\Delta n_1
+2t_1t_2\gamma^{ke}_{13}}{T_1}
+\frac{-2t^2\Delta n_1
-2t^2\gamma_{13}}{T_1}
-\frac{2Ut}{T_1}\sum_{\sigma\neq \sigma'}\langle \Psi|\hat{\gamma}_{1,2}^\sigma (\hat{n}_2^{\sigma'}-\hat{n}_1^{\sigma'})|\Psi\rangle
\label{eq:v^Mxc_1_4sites}
\end{eqnarray}
\begin{align}
\Delta v_2^{Mxc}
[\mathbf{ n}]&=\frac{2t_2^2\Delta n_2-2t_1t_2\gamma^{ke}_{13}+2t_2t_3\gamma^{ke}_{24}}{T_2}
+\frac{-2t^2\Delta n_2
+2t^2\gamma_{13}
-t^2\gamma_{24}}{T_2}\nonumber\\
&-\frac{2Ut}{T_2}\sum_{\sigma\neq \sigma'}\langle \Psi|\hat{\gamma}_{2,3}^\sigma (\hat{n}_3^{\sigma'}-\hat{n}_3^{\sigma'})|\Psi\rangle
\label{eq:v^Mxc_2_4sites}
ns \end{align}

\begin{eqnarray}
\Delta v_3^{Mxc}
[\mathbf{ n}]=\frac{2t_3^2\Delta n_3
-2t_2t_3\gamma^{ke}_{24}}{T_3}+
\frac{-2t^2\Delta n_3+2t^2\gamma_{24}}{T_3}
-\frac{2Ut}{T_3}\sum_{\sigma\neq \sigma'}\langle \Psi|\hat{\gamma}_{3,4}^\sigma (\hat{n}_4^{\sigma'}-\hat{n}_3^{\sigma'})|\Psi\rangle
\label{eq:v^Mxc_3_4sites}
\end{eqnarray}
Similary to $\mathbf{v}^{Hxc}$ we decompose $\mathbf{v}^{Mxc}$ in a $\mathbf{v}^{Mx}$ and a correlation part $\mathbf{v}^{c}$. This analogous to the decomposition for the two-site case presented in Sec.\ \ref{sec:Results}:
\begin{eqnarray}
\Delta v_1^{Mx}
[\mathbf{ n},\Phi]=\frac{2t_1^2\Delta n_1+
2t_1t_2\gamma^{ke}_{13}}{T_1}
+\frac{-2t^2\Delta n_1
-2t^2\gamma_{13}^{ke}}{T_1}
-\frac{2Ut}{T_1}\sum_{\sigma\neq \sigma'}\langle \Phi|\hat{\gamma}_{1,2}^\sigma (\hat{n}_2^{\sigma'}-\hat{n}_1^{\sigma'})|\Phi\rangle
\label{eq:v^Mx_1_4sites}
\end{eqnarray}
\begin{align}
\Delta v_2^{Mx}
[\mathbf{ n},\Phi]&=\frac{2t_2^2\Delta n_2-2t_1t_2\gamma^{ke}_{13}+2t_2t_3\gamma^{ke}_{24}}{T_2}
+\frac{-2t^2\Delta n_2
+2t^2\gamma_{13}^{ke}
-2t^2\gamma_{24}^{ke}}{T_2}\nonumber\\
&-\frac{2Ut}{T_2}\sum_{\sigma\neq \sigma'}\langle \Phi|\hat{\gamma}_{2,3}^\sigma (\hat{n}_3^{\sigma'}-\hat{n}_3^{\sigma'})|\Phi\rangle
\label{eq:v^Mx_2_4sites}
\end{align}
\begin{align}
\Delta v_3^{Mx}
[\mathbf{ n}]=\frac{2t_3^2\Delta n_3
-2t_2t_3\gamma^{ke}_{24}}{T_3}+
\frac{-2t^2\Delta n_3+2t^2\gamma^{ke}_{24}}{T_3}
-\frac{2Ut}{T_3}\sum_{\sigma\neq \sigma'}\langle \Phi|\hat{\gamma}_{3,4}^\sigma (\hat{n}_4^{\sigma'}-\hat{n}_3^{\sigma'})|\Phi\rangle
\label{eq:v^Mx_3_4sites}
\end{align}
\begin{align}
\Delta v_{c,1}^{ke}[\mathbf{ n},\Phi]&=
\frac{2t^2\gamma^{ke}_{13}}{T_1}
-\frac{2t^2\gamma_{13}}{T_1}
\nonumber\\
&-\frac{2Ut}{T_1}\sum_{
\sigma \neq \sigma'}\langle \Psi |\hat{\gamma}_{1,2}^\sigma (\hat{n}_2^{\sigma'}-\hat{n}_1^{\sigma'})|\Psi\rangle+\frac{2Ut}{T_1}\sum_{
\sigma \neq \sigma'}\langle \Phi |\hat{\gamma}_{1,2}^\sigma(\hat{n}_2^{\sigma'}-\hat{n}_1^{\sigma'})|\Phi\rangle
\label{eq:vc_1_ke_4sites}
\end{align}

\begin{align}
\Delta v_{c,2}^{ke}[\mathbf{ n},\Phi]&=
\frac{-2t^2\gamma^{ke}_{13}+2t^2\gamma_{24}^{ke}}{T_2}
+\frac{2t^2\gamma_{13}
-2t^2\gamma_{24}}{T_2}
\nonumber\\
&-\frac{2Ut}{T_2}\sum_{
\sigma \neq \sigma'}\langle \Psi |\hat{\gamma}_{2,3}^\sigma (\hat{n}_3^{\sigma'}-\hat{n}_2^{\sigma'})|\Psi\rangle
+\frac{2Ut}{T_2}\sum_{
\sigma \neq \sigma'}\langle \Phi |\hat{\gamma}_{2,3}^\sigma(\hat{n}_3^{\sigma'}-\hat{n}_2^{\sigma'})|\Phi\rangle
\label{eq:vc_2^ke_4sites}
\end{align}

\begin{align}
\Delta v_{c,3}^{ke}
[\mathbf{ n}]&=
\frac{-2t^2\gamma^{ke}_{24}}{T_3}
+\frac{2t^2\gamma_{24}}{T_3}\nonumber\\
&-\frac{2Ut}{T_3}\sum_{\sigma\neq \sigma'}\langle \Psi|\hat{\gamma}_{3,4}^\sigma (\hat{n}_4^{\sigma'}-\hat{n}_3^{\sigma'})|\Psi\rangle
+\frac{2Ut}{T_3}\sum_{\sigma\neq \sigma'}\langle \Phi|\hat{\gamma}_{3,4}^\sigma (\hat{n}_4^{\sigma'}-\hat{n}_3^{\sigma'})|\Phi\rangle
\label{eq:vc_3_ke_4sites}
\end{align}

\par
\section{Gauge choice for hopping}
\label{app:Gauge}
In this Appendix, we demonstrate that the sign of the hopping is just a gauge choice. Let us consider a single-particle Hamiltonian $H$. In the site basis, it corresponds to the Matrix
\begin{align}
  \mat{H} & = \begin{pmatrix}
  h_{11} & h_{12} & \cdots \\
  h_{12}^\star & h_{22} & \cdots \\
  \vdots & \vdots & \ddots
  \end{pmatrix} = \mat{U} \cdot \begin{pmatrix}
  \epsilon_1 & 0 & \cdots \\
  0 & \epsilon_2 & \cdots \\
  \vdots & \vdots & \ddots
  \end{pmatrix} \cdot \mat{U}^\dagger ~, \label{H}
\end{align}
where we also introduced its spectral decomposition with the unitary matrix $\mat{U}$ having its eigenvectors as columns. Suppose we transform the Hamiltonian by 
\begin{align}
  \mat{G} = \begin{pmatrix}
  \ID_k & 0 \\
  0 & -\ID_{N-k} 
  \end{pmatrix} ~, \label{G}
\end{align}
where $\ID_n$ is the $n\times n$ unit matrix and $N$ is the total number of sites. Note that $\mat{G}^\dagger = \mat{G}^{-1}$, i.e., $\mat{G}$ is unitary. It is straight forward to verify that the transformed Hamiltonian is given by
\begin{align}
  &\mat{H}' = \mat{G} \cdot \mat{H} \cdot \mat{G}^\dagger 
  = \\
  &\begin{pmatrix}
  h_{11} & \cdots & h_{1 k} & -h_{1 (k+1)} & \cdots & - h_{1 N} \\
  \vdots & \ddots & \vdots  & \vdots & \ddots & \vdots \\
  h_{k1} & \cdots & h_{k k} & -h_{k (k+1)} & \cdots & - h_{k N} \\
  -h_{(k+1) 1} & \cdots & -h_{(k+1) k} & h_{k+1 k+1} & \cdots & h_{(k+1) N} \\
  \vdots & \ddots & \vdots  & \vdots & \ddots & \vdots \\
  -h_{N 1} & \cdots & -h_{N k} & h_{N (k+1)} & \cdots & h_{N N}
  \end{pmatrix} ~. \nonumber\label{Hprime}
\end{align}
We can see that the effect of the transformation is to change the sign of all matrix elements in the ``off-diagonal'' parts determined by the site $k$ after which the sign in $\mat{G}$ changes. For a \emph{one-dimensional} nearest-neighbor tight-binding Hamiltonian with ``zero boundary conditions'', like the one we consider in this manuscript, this corresponds to flipping the sign of the hopping amplitude between site $k$ and $k+1$. Obviously the eigenvalues of $\mat{H}'$ are the same as the eigenvalues of $\mat{H}$, and the eigenstates are simply given by $\mat{U}' = \mat{G} \mat{U}$, which means that the signs of the wave function in position representation are flipped from site $k+1$ onwards. All Hamiltonians, which can be connected by such a transformation are to be considered equivalent.  

\end{appendix}

\bibliography{kin_energy}

\begin{thebibliography}{56}%
\makeatletter
\providecommand \@ifxundefined [1]{%
 \@ifx{#1\undefined}
}%
\providecommand \@ifnum [1]{%
 \ifnum #1\expandafter \@firstoftwo
 \else \expandafter \@secondoftwo
 \fi
}%
\providecommand \@ifx [1]{%
 \ifx #1\expandafter \@firstoftwo
 \else \expandafter \@secondoftwo
 \fi
}%
\providecommand \natexlab [1]{#1}%
\providecommand \enquote  [1]{``#1''}%
\providecommand \bibnamefont  [1]{#1}%
\providecommand \bibfnamefont [1]{#1}%
\providecommand \citenamefont [1]{#1}%
\providecommand \href@noop [0]{\@secondoftwo}%
\providecommand \href [0]{\begingroup \@sanitize@url \@href}%
\providecommand \@href[1]{\@@startlink{#1}\@@href}%
\providecommand \@@href[1]{\endgroup#1\@@endlink}%
\providecommand \@sanitize@url [0]{\catcode `\\12\catcode `\$12\catcode
  `\&12\catcode `\#12\catcode `\^12\catcode `\_12\catcode `\%12\relax}%
\providecommand \@@startlink[1]{}%
\providecommand \@@endlink[0]{}%
\providecommand \url  [0]{\begingroup\@sanitize@url \@url }%
\providecommand \@url [1]{\endgroup\@href {#1}{\urlprefix }}%
\providecommand \urlprefix  [0]{URL }%
\providecommand \Eprint [0]{\href }%
\providecommand \doibase [0]{http://dx.doi.org/}%
\providecommand \selectlanguage [0]{\@gobble}%
\providecommand \bibinfo  [0]{\@secondoftwo}%
\providecommand \bibfield  [0]{\@secondoftwo}%
\providecommand \translation [1]{[#1]}%
\providecommand \BibitemOpen [0]{}%
\providecommand \bibitemStop [0]{}%
\providecommand \bibitemNoStop [0]{.\EOS\space}%
\providecommand \EOS [0]{\spacefactor3000\relax}%
\providecommand \BibitemShut  [1]{\csname bibitem#1\endcsname}%
\let\auto@bib@innerbib\@empty
\bibitem [{\citenamefont {Hohenberg}\ and\ \citenamefont
  {Kohn}(1964)}]{Hohenberg-Kohn}%
  \BibitemOpen
  \bibfield  {author} {\bibinfo {author} {\bibfnamefont {P.}~\bibnamefont
  {Hohenberg}}\ and\ \bibinfo {author} {\bibfnamefont {W.}~\bibnamefont
  {Kohn}},\ }\href {\doibase 10.1103/PhysRev.136.B864} {\bibfield  {journal}
  {\bibinfo  {journal} {Phys. Rev.}\ }\textbf {\bibinfo {volume} {136}},\
  \bibinfo {pages} {B864} (\bibinfo {year} {1964})}\BibitemShut {NoStop}%
\bibitem [{\citenamefont {Thomas}(1927)}]{thomas_1927}%
  \BibitemOpen
  \bibfield  {author} {\bibinfo {author} {\bibfnamefont {L.~H.}\ \bibnamefont
  {Thomas}},\ }\href {\doibase 10.1017/S0305004100011683} {\bibfield  {journal}
  {\bibinfo  {journal} {Mathematical Proceedings of the Cambridge Philosophical
  Society}\ }\textbf {\bibinfo {volume} {23}},\ \bibinfo {pages} {542–548}
  (\bibinfo {year} {1927})}\BibitemShut {NoStop}%
\bibitem [{\citenamefont {Fermi}(1928)}]{Fermi1928}%
  \BibitemOpen
  \bibfield  {author} {\bibinfo {author} {\bibfnamefont {E.}~\bibnamefont
  {Fermi}},\ }\href {\doibase 10.1007/BF01351576} {\bibfield  {journal}
  {\bibinfo  {journal} {Zeitschrift f{\"u}r Physik}\ }\textbf {\bibinfo
  {volume} {48}},\ \bibinfo {pages} {73} (\bibinfo {year} {1928})}\BibitemShut
  {NoStop}%
\bibitem [{\citenamefont {Lieb}(1976)}]{Stability_of_matter_thomas_fermi}%
  \BibitemOpen
  \bibfield  {author} {\bibinfo {author} {\bibfnamefont {E.~H.}\ \bibnamefont
  {Lieb}},\ }\href {\doibase 10.1103/RevModPhys.48.553} {\bibfield  {journal}
  {\bibinfo  {journal} {Rev. Mod. Phys.}\ }\textbf {\bibinfo {volume} {48}},\
  \bibinfo {pages} {553} (\bibinfo {year} {1976})}\BibitemShut {NoStop}%
\bibitem [{\citenamefont {Kohn}\ and\ \citenamefont {Sham}(1965)}]{Kohn-Sham}%
  \BibitemOpen
  \bibfield  {author} {\bibinfo {author} {\bibfnamefont {W.}~\bibnamefont
  {Kohn}}\ and\ \bibinfo {author} {\bibfnamefont {L.~J.}\ \bibnamefont
  {Sham}},\ }\href {\doibase 10.1103/PhysRev.140.A1133} {\bibfield  {journal}
  {\bibinfo  {journal} {Phys. Rev.}\ }\textbf {\bibinfo {volume} {140}},\
  \bibinfo {pages} {A1133} (\bibinfo {year} {1965})}\BibitemShut {NoStop}%
\bibitem [{\citenamefont {Burke}(2012)}]{Burke_DFT_review}%
  \BibitemOpen
  \bibfield  {author} {\bibinfo {author} {\bibfnamefont {K.}~\bibnamefont
  {Burke}},\ }\href {\doibase 10.1063/1.4704546} {\bibfield  {journal}
  {\bibinfo  {journal} {The Journal of Chemical Physics}\ }\textbf {\bibinfo
  {volume} {136}},\ \bibinfo {pages} {150901} (\bibinfo {year} {2012})},\
  \Eprint {http://arxiv.org/abs/https://doi.org/10.1063/1.4704546}
  {https://doi.org/10.1063/1.4704546} \BibitemShut {NoStop}%
\bibitem [{\citenamefont {Medvedev}\ \emph {et~al.}(2017)\citenamefont
  {Medvedev}, \citenamefont {Bushmarinov}, \citenamefont {Sun}, \citenamefont
  {Perdew},\ and\ \citenamefont {Lyssenko}}]{Science_Perdew}%
  \BibitemOpen
  \bibfield  {author} {\bibinfo {author} {\bibfnamefont {M.~G.}\ \bibnamefont
  {Medvedev}}, \bibinfo {author} {\bibfnamefont {I.~S.}\ \bibnamefont
  {Bushmarinov}}, \bibinfo {author} {\bibfnamefont {J.}~\bibnamefont {Sun}},
  \bibinfo {author} {\bibfnamefont {J.~P.}\ \bibnamefont {Perdew}}, \ and\
  \bibinfo {author} {\bibfnamefont {K.~A.}\ \bibnamefont {Lyssenko}},\ }\href
  {\doibase 10.1126/science.aah5975} {\bibfield  {journal} {\bibinfo  {journal}
  {Science}\ }\textbf {\bibinfo {volume} {355}},\ \bibinfo {pages} {49}
  (\bibinfo {year} {2017})},\ \Eprint
  {http://arxiv.org/abs/http://science.sciencemag.org/content/355/6320/49.full.pdf}
  {http://science.sciencemag.org/content/355/6320/49.full.pdf} \BibitemShut
  {NoStop}%
\bibitem [{\citenamefont {Alexander L.~Fetter}(1971)}]{Fetter_Walecka}%
  \BibitemOpen
  \bibfield  {author} {\bibinfo {author} {\bibfnamefont {J.~D.~W.}\
  \bibnamefont {Alexander L.~Fetter}},\ }\href@noop {} {\emph {\bibinfo {title}
  {Quantum theory of many-particle systems}}}\ (\bibinfo  {publisher}
  {McGraw-Hill Book Company},\ \bibinfo {address} {New York},\ \bibinfo {year}
  {1971})\BibitemShut {NoStop}%
\bibitem [{\citenamefont {Gianluca~Stefanucci}(2013)}]{Stefanucci_Van_Leeuwen}%
  \BibitemOpen
  \bibfield  {author} {\bibinfo {author} {\bibfnamefont {R.~v.~L.}\
  \bibnamefont {Gianluca~Stefanucci}},\ }\href@noop {} {\emph {\bibinfo {title}
  {Nonequilibrium Many-Body Theory of Quantum Systems}}}\ (\bibinfo
  {publisher} {Cambridge University Press},\ \bibinfo {address} {Cambridge},\
  \bibinfo {year} {2013})\BibitemShut {NoStop}%
\bibitem [{\citenamefont {Mazziotti}(2006)}]{mazziotti_review}%
  \BibitemOpen
  \bibfield  {author} {\bibinfo {author} {\bibfnamefont {D.~A.}\ \bibnamefont
  {Mazziotti}},\ }\href {\doibase 10.1021/ar050029d} {\bibfield  {journal}
  {\bibinfo  {journal} {Accounts of Chemical Research}\ }\textbf {\bibinfo
  {volume} {39}},\ \bibinfo {pages} {207} (\bibinfo {year} {2006})}\BibitemShut
  {NoStop}%
\bibitem [{\citenamefont {Pernal}\ and\ \citenamefont
  {Giesbertz}(2016)}]{Review-RDMFT_Katarzyna_Klaas}%
  \BibitemOpen
  \bibfield  {author} {\bibinfo {author} {\bibfnamefont {K.}~\bibnamefont
  {Pernal}}\ and\ \bibinfo {author} {\bibfnamefont {K.~J.~H.}\ \bibnamefont
  {Giesbertz}},\ }\enquote {\bibinfo {title} {Reduced density matrix functional
  theory (rdmft) and linear response time-dependent rdmft (td-rdmft)},}\ in\
  \href {\doibase 10.1007/128_2015_624} {\emph {\bibinfo {booktitle}
  {Density-Functional Methods for Excited States}}},\ \bibinfo {editor} {edited
  by\ \bibinfo {editor} {\bibfnamefont {N.}~\bibnamefont {Ferr{\'e}}}, \bibinfo
  {editor} {\bibfnamefont {M.}~\bibnamefont {Filatov}}, \ and\ \bibinfo
  {editor} {\bibfnamefont {M.}~\bibnamefont {Huix-Rotllant}}}\ (\bibinfo
  {publisher} {Springer International Publishing},\ \bibinfo {address} {Cham},\
  \bibinfo {year} {2016})\ pp.\ \bibinfo {pages} {125--183}\BibitemShut
  {NoStop}%
\bibitem [{\citenamefont {Mazziotti}(2012)}]{Nrep-2RDM}%
  \BibitemOpen
  \bibfield  {author} {\bibinfo {author} {\bibfnamefont {D.~A.}\ \bibnamefont
  {Mazziotti}},\ }\href {\doibase 10.1103/PhysRevLett.108.263002} {\bibfield
  {journal} {\bibinfo  {journal} {Phys. Rev. Lett.}\ }\textbf {\bibinfo
  {volume} {108}},\ \bibinfo {pages} {263002} (\bibinfo {year}
  {2012})}\BibitemShut {NoStop}%
\bibitem [{\citenamefont {Klyachko}(2006)}]{Klyachko_GPC}%
  \BibitemOpen
  \bibfield  {author} {\bibinfo {author} {\bibfnamefont {A.~A.}\ \bibnamefont
  {Klyachko}},\ }\href@noop {} {\bibfield  {journal} {\bibinfo  {journal}
  {Journal of Physics: Conference Series}\ }\textbf {\bibinfo {volume} {36}},\
  \bibinfo {pages} {72} (\bibinfo {year} {2006})}\BibitemShut {NoStop}%
\bibitem [{\citenamefont {Bonitz}(2016)}]{Bonitz}%
  \BibitemOpen
  \bibfield  {author} {\bibinfo {author} {\bibfnamefont {M.}~\bibnamefont
  {Bonitz}},\ }\href@noop {} {\emph {\bibinfo {title} {Quantum Kinetic
  Theory}}},\ \bibinfo {edition} {2nd}\ ed.\ (\bibinfo  {publisher}
  {Springer},\ \bibinfo {year} {2016})\BibitemShut {NoStop}%
\bibitem [{\citenamefont {Baldsiefen}\ \emph {et~al.}(2015)\citenamefont
  {Baldsiefen}, \citenamefont {Cangi},\ and\ \citenamefont
  {Gross}}]{baldsiefen_reduced_thermal}%
  \BibitemOpen
  \bibfield  {author} {\bibinfo {author} {\bibfnamefont {T.}~\bibnamefont
  {Baldsiefen}}, \bibinfo {author} {\bibfnamefont {A.}~\bibnamefont {Cangi}}, \
  and\ \bibinfo {author} {\bibfnamefont {E.}~\bibnamefont {Gross}},\
  }\href@noop {} {\bibfield  {journal} {\bibinfo  {journal} {Physical Review
  A}\ }\textbf {\bibinfo {volume} {92}},\ \bibinfo {pages} {052514} (\bibinfo
  {year} {2015})}\BibitemShut {NoStop}%
\bibitem [{\citenamefont {{Giesbertz}}\ and\ \citenamefont
  {{Ruggenthaler}}(2017)}]{RDM_elevated_temperature}%
  \BibitemOpen
  \bibfield  {author} {\bibinfo {author} {\bibfnamefont {K.~J.~H.}\
  \bibnamefont {{Giesbertz}}}\ and\ \bibinfo {author} {\bibfnamefont
  {M.}~\bibnamefont {{Ruggenthaler}}},\ }\href@noop {} {\bibfield  {journal}
  {\bibinfo  {journal} {ArXiv e-prints}\ } (\bibinfo {year} {2017})},\ \Eprint
  {http://arxiv.org/abs/1710.08805} {arXiv:1710.08805 [math-ph]} \BibitemShut
  {NoStop}%
\bibitem [{\citenamefont {Lathiotakis}\ \emph {et~al.}(2014)\citenamefont
  {Lathiotakis}, \citenamefont {Helbig}, \citenamefont {Rubio},\ and\
  \citenamefont {Gidopoulos}}]{Local_RDMFT}%
  \BibitemOpen
  \bibfield  {author} {\bibinfo {author} {\bibfnamefont {N.~N.}\ \bibnamefont
  {Lathiotakis}}, \bibinfo {author} {\bibfnamefont {N.}~\bibnamefont {Helbig}},
  \bibinfo {author} {\bibfnamefont {A.}~\bibnamefont {Rubio}}, \ and\ \bibinfo
  {author} {\bibfnamefont {N.~I.}\ \bibnamefont {Gidopoulos}},\ }\href
  {\doibase 10.1103/PhysRevA.90.032511} {\bibfield  {journal} {\bibinfo
  {journal} {Phys. Rev. A}\ }\textbf {\bibinfo {volume} {90}},\ \bibinfo
  {pages} {032511} (\bibinfo {year} {2014})}\BibitemShut {NoStop}%
\bibitem [{\citenamefont {Gori-Giorgi}\ \emph {et~al.}(2009)\citenamefont
  {Gori-Giorgi}, \citenamefont {Seidl},\ and\ \citenamefont
  {Vignale}}]{gori-giorgi2009}%
  \BibitemOpen
  \bibfield  {author} {\bibinfo {author} {\bibfnamefont {P.}~\bibnamefont
  {Gori-Giorgi}}, \bibinfo {author} {\bibfnamefont {M.}~\bibnamefont {Seidl}},
  \ and\ \bibinfo {author} {\bibfnamefont {G.}~\bibnamefont {Vignale}},\ }\href
  {\doibase 10.1103/PhysRevLett.103.166402} {\bibfield  {journal} {\bibinfo
  {journal} {Phys. Rev. Lett.}\ }\textbf {\bibinfo {volume} {103}},\ \bibinfo
  {pages} {166402} (\bibinfo {year} {2009})}\BibitemShut {NoStop}%
\bibitem [{\citenamefont {Malet}\ \emph {et~al.}(2014)\citenamefont {Malet},
  \citenamefont {Mirtschink}, \citenamefont {Giesbertz}, \citenamefont
  {Wagner},\ and\ \citenamefont {Gori-Giorgi}}]{malet2014exchange}%
  \BibitemOpen
  \bibfield  {author} {\bibinfo {author} {\bibfnamefont {F.}~\bibnamefont
  {Malet}}, \bibinfo {author} {\bibfnamefont {A.}~\bibnamefont {Mirtschink}},
  \bibinfo {author} {\bibfnamefont {K.~J.}\ \bibnamefont {Giesbertz}}, \bibinfo
  {author} {\bibfnamefont {L.~O.}\ \bibnamefont {Wagner}}, \ and\ \bibinfo
  {author} {\bibfnamefont {P.}~\bibnamefont {Gori-Giorgi}},\ }\href@noop {}
  {\bibfield  {journal} {\bibinfo  {journal} {Physical Chemistry Chemical
  Physics}\ }\textbf {\bibinfo {volume} {16}},\ \bibinfo {pages} {14551}
  (\bibinfo {year} {2014})}\BibitemShut {NoStop}%
\bibitem [{\citenamefont {Grossi}\ \emph {et~al.}(2017)\citenamefont {Grossi},
  \citenamefont {Kooi}, \citenamefont {Giesbertz}, \citenamefont {Seidl},
  \citenamefont {Cohen}, \citenamefont {Mori-S{\'a}nchez},\ and\ \citenamefont
  {Gori-Giorgi}}]{grossi2017fermionic}%
  \BibitemOpen
  \bibfield  {author} {\bibinfo {author} {\bibfnamefont {J.}~\bibnamefont
  {Grossi}}, \bibinfo {author} {\bibfnamefont {D.~P.}\ \bibnamefont {Kooi}},
  \bibinfo {author} {\bibfnamefont {K.~J.}\ \bibnamefont {Giesbertz}}, \bibinfo
  {author} {\bibfnamefont {M.}~\bibnamefont {Seidl}}, \bibinfo {author}
  {\bibfnamefont {A.~J.}\ \bibnamefont {Cohen}}, \bibinfo {author}
  {\bibfnamefont {P.}~\bibnamefont {Mori-S{\'a}nchez}}, \ and\ \bibinfo
  {author} {\bibfnamefont {P.}~\bibnamefont {Gori-Giorgi}},\ }\href@noop {}
  {\bibfield  {journal} {\bibinfo  {journal} {Journal of chemical theory and
  computation}\ }\textbf {\bibinfo {volume} {13}},\ \bibinfo {pages} {6089}
  (\bibinfo {year} {2017})}\BibitemShut {NoStop}%
\bibitem [{\citenamefont {Eich}\ \emph {et~al.}(2014)\citenamefont {Eich},
  \citenamefont {Di~Ventra},\ and\ \citenamefont {Vignale}}]{ThermalDFT_PRL}%
  \BibitemOpen
  \bibfield  {author} {\bibinfo {author} {\bibfnamefont {F.~G.}\ \bibnamefont
  {Eich}}, \bibinfo {author} {\bibfnamefont {M.}~\bibnamefont {Di~Ventra}}, \
  and\ \bibinfo {author} {\bibfnamefont {G.}~\bibnamefont {Vignale}},\ }\href
  {\doibase 10.1103/PhysRevLett.112.196401} {\bibfield  {journal} {\bibinfo
  {journal} {Phys. Rev. Lett.}\ }\textbf {\bibinfo {volume} {112}},\ \bibinfo
  {pages} {196401} (\bibinfo {year} {2014})}\BibitemShut {NoStop}%
\bibitem [{\citenamefont {Eich}\ \emph {et~al.}(2017)\citenamefont {Eich},
  \citenamefont {Ventra},\ and\ \citenamefont {Vignale}}]{thermal_DFT}%
  \BibitemOpen
  \bibfield  {author} {\bibinfo {author} {\bibfnamefont {F.~G.}\ \bibnamefont
  {Eich}}, \bibinfo {author} {\bibfnamefont {M.~D.}\ \bibnamefont {Ventra}}, \
  and\ \bibinfo {author} {\bibfnamefont {G.}~\bibnamefont {Vignale}},\ }\href
  {http://stacks.iop.org/0953-8984/29/i=6/a=063001} {\bibfield  {journal}
  {\bibinfo  {journal} {Journal of Physics: Condensed Matter}\ }\textbf
  {\bibinfo {volume} {29}},\ \bibinfo {pages} {063001} (\bibinfo {year}
  {2017})}\BibitemShut {NoStop}%
\bibitem [{\citenamefont {Ghosh}\ \emph {et~al.}(1984)\citenamefont {Ghosh},
  \citenamefont {Berkowitz},\ and\ \citenamefont
  {Parr}}]{Ghosh_DFT_thermodynamics}%
  \BibitemOpen
  \bibfield  {author} {\bibinfo {author} {\bibfnamefont {S.~K.}\ \bibnamefont
  {Ghosh}}, \bibinfo {author} {\bibfnamefont {M.}~\bibnamefont {Berkowitz}}, \
  and\ \bibinfo {author} {\bibfnamefont {R.~G.}\ \bibnamefont {Parr}},\
  }\href@noop {} {\bibfield  {journal} {\bibinfo  {journal} {Proceedings of the
  National Academy of Sciences}\ }\textbf {\bibinfo {volume} {81}},\ \bibinfo
  {pages} {8028} (\bibinfo {year} {1984})}\BibitemShut {NoStop}%
\bibitem [{\citenamefont {Ayers}\ \emph {et~al.}(2002)\citenamefont {Ayers},
  \citenamefont {Parr},\ and\ \citenamefont
  {Nagy}}]{Ayers_Parr_Nagy_local_temperature}%
  \BibitemOpen
  \bibfield  {author} {\bibinfo {author} {\bibfnamefont {P.~W.}\ \bibnamefont
  {Ayers}}, \bibinfo {author} {\bibfnamefont {R.~G.}\ \bibnamefont {Parr}}, \
  and\ \bibinfo {author} {\bibfnamefont {A.}~\bibnamefont {Nagy}},\ }\href
  {\doibase 10.1002/qua.989} {\bibfield  {journal} {\bibinfo  {journal}
  {International Journal of Quantum Chemistry}\ }\textbf {\bibinfo {volume}
  {90}},\ \bibinfo {pages} {309} (\bibinfo {year} {2002})}\BibitemShut
  {NoStop}%
\bibitem [{\citenamefont {Nagy}(2017)}]{Nagy_dft_thermodynamics}%
  \BibitemOpen
  \bibfield  {author} {\bibinfo {author} {\bibfnamefont {Ã.}~\bibnamefont
  {Nagy}},\ }\href {\doibase 10.1002/qua.25396} {\bibfield  {journal} {\bibinfo
   {journal} {International Journal of Quantum Chemistry}\ }\textbf {\bibinfo
  {volume} {117}},\ \bibinfo {pages} {e25396} (\bibinfo {year} {2017})},\
  \bibinfo {note} {e25396}\BibitemShut {NoStop}%
\bibitem [{\citenamefont {Seidl}\ \emph {et~al.}(1996)\citenamefont {Seidl},
  \citenamefont {G\"orling}, \citenamefont {Vogl}, \citenamefont {Majewski},\
  and\ \citenamefont {Levy}}]{SeidlLevy:96}%
  \BibitemOpen
  \bibfield  {author} {\bibinfo {author} {\bibfnamefont {A.}~\bibnamefont
  {Seidl}}, \bibinfo {author} {\bibfnamefont {A.}~\bibnamefont {G\"orling}},
  \bibinfo {author} {\bibfnamefont {P.}~\bibnamefont {Vogl}}, \bibinfo {author}
  {\bibfnamefont {J.~A.}\ \bibnamefont {Majewski}}, \ and\ \bibinfo {author}
  {\bibfnamefont {M.}~\bibnamefont {Levy}},\ }\href {\doibase
  10.1103/PhysRevB.53.3764} {\bibfield  {journal} {\bibinfo  {journal} {Phys.
  Rev. B}\ }\textbf {\bibinfo {volume} {53}},\ \bibinfo {pages} {3764}
  (\bibinfo {year} {1996})}\BibitemShut {NoStop}%
\bibitem [{\citenamefont {Eich}\ and\ \citenamefont
  {Hellgren}(2014)}]{EichHellgren:14}%
  \BibitemOpen
  \bibfield  {author} {\bibinfo {author} {\bibfnamefont {F.~G.}\ \bibnamefont
  {Eich}}\ and\ \bibinfo {author} {\bibfnamefont {M.}~\bibnamefont
  {Hellgren}},\ }\href {\doibase http://dx.doi.org/10.1063/1.4903273}
  {\bibfield  {journal} {\bibinfo  {journal} {The Journal of Chemical Physics}\
  }\textbf {\bibinfo {volume} {141}},\ \bibinfo {eid} {224107} (\bibinfo {year}
  {2014})}\BibitemShut {NoStop}%
\bibitem [{\citenamefont {Dimitrov}\ \emph {et~al.}(2016)\citenamefont
  {Dimitrov}, \citenamefont {Appel}, \citenamefont {Fuks},\ and\ \citenamefont
  {Rubio}}]{Tanja_exact_maps_lattice}%
  \BibitemOpen
  \bibfield  {author} {\bibinfo {author} {\bibfnamefont {T.}~\bibnamefont
  {Dimitrov}}, \bibinfo {author} {\bibfnamefont {H.}~\bibnamefont {Appel}},
  \bibinfo {author} {\bibfnamefont {J.~I.}\ \bibnamefont {Fuks}}, \ and\
  \bibinfo {author} {\bibfnamefont {A.}~\bibnamefont {Rubio}},\ }\href@noop {}
  {\bibfield  {journal} {\bibinfo  {journal} {New Journal of Physics}\ }\textbf
  {\bibinfo {volume} {18}},\ \bibinfo {pages} {083004} (\bibinfo {year}
  {2016})}\BibitemShut {NoStop}%
\bibitem [{\citenamefont {Carrascal}\ \emph {et~al.}(2015)\citenamefont
  {Carrascal}, \citenamefont {Ferrer}, \citenamefont {Smith},\ and\
  \citenamefont {Burke}}]{2site_Hubbard_Burke}%
  \BibitemOpen
  \bibfield  {author} {\bibinfo {author} {\bibfnamefont {D.~J.}\ \bibnamefont
  {Carrascal}}, \bibinfo {author} {\bibfnamefont {J.}~\bibnamefont {Ferrer}},
  \bibinfo {author} {\bibfnamefont {J.~C.}\ \bibnamefont {Smith}}, \ and\
  \bibinfo {author} {\bibfnamefont {K.}~\bibnamefont {Burke}},\ }\href
  {http://stacks.iop.org/0953-8984/27/i=39/a=393001} {\bibfield  {journal}
  {\bibinfo  {journal} {Journal of Physics: Condensed Matter}\ }\textbf
  {\bibinfo {volume} {27}},\ \bibinfo {pages} {393001} (\bibinfo {year}
  {2015})}\BibitemShut {NoStop}%
\bibitem [{\citenamefont {Cohen}\ and\ \citenamefont
  {Mori-S\'anchez}(2016)}]{Exact_DFT_functional_2sites}%
  \BibitemOpen
  \bibfield  {author} {\bibinfo {author} {\bibfnamefont {A.~J.}\ \bibnamefont
  {Cohen}}\ and\ \bibinfo {author} {\bibfnamefont {P.}~\bibnamefont
  {Mori-S\'anchez}},\ }\href {\doibase 10.1103/PhysRevA.93.042511} {\bibfield
  {journal} {\bibinfo  {journal} {Phys. Rev. A}\ }\textbf {\bibinfo {volume}
  {93}},\ \bibinfo {pages} {042511} (\bibinfo {year} {2016})}\BibitemShut
  {NoStop}%
\bibitem [{\citenamefont {Dimitrov}\ \emph {et~al.}(2017)\citenamefont
  {Dimitrov}, \citenamefont {Flick}, \citenamefont {Ruggenthaler},\ and\
  \citenamefont {Rubio}}]{Tanja_exact_photon_maps}%
  \BibitemOpen
  \bibfield  {author} {\bibinfo {author} {\bibfnamefont {T.}~\bibnamefont
  {Dimitrov}}, \bibinfo {author} {\bibfnamefont {J.}~\bibnamefont {Flick}},
  \bibinfo {author} {\bibfnamefont {M.}~\bibnamefont {Ruggenthaler}}, \ and\
  \bibinfo {author} {\bibfnamefont {A.}~\bibnamefont {Rubio}},\ }\href
  {http://stacks.iop.org/1367-2630/19/i=11/a=113036} {\bibfield  {journal}
  {\bibinfo  {journal} {New Journal of Physics}\ }\textbf {\bibinfo {volume}
  {19}},\ \bibinfo {pages} {113036} (\bibinfo {year} {2017})}\BibitemShut
  {NoStop}%
\bibitem [{\citenamefont {Lammert}(2010)}]{coarse_grained_DFT}%
  \BibitemOpen
  \bibfield  {author} {\bibinfo {author} {\bibfnamefont {P.~E.}\ \bibnamefont
  {Lammert}},\ }\href {\doibase 10.1103/PhysRevA.82.012109} {\bibfield
  {journal} {\bibinfo  {journal} {Phys. Rev. A}\ }\textbf {\bibinfo {volume}
  {82}},\ \bibinfo {pages} {012109} (\bibinfo {year} {2010})}\BibitemShut
  {NoStop}%
\bibitem [{\citenamefont {Kvaal}\ \emph {et~al.}(2014)\citenamefont {Kvaal},
  \citenamefont {Ekström}, \citenamefont {Teale},\ and\ \citenamefont
  {Helgaker}}]{differentiable_DFT}%
  \BibitemOpen
  \bibfield  {author} {\bibinfo {author} {\bibfnamefont {S.}~\bibnamefont
  {Kvaal}}, \bibinfo {author} {\bibfnamefont {U.}~\bibnamefont {Ekström}},
  \bibinfo {author} {\bibfnamefont {A.~M.}\ \bibnamefont {Teale}}, \ and\
  \bibinfo {author} {\bibfnamefont {T.}~\bibnamefont {Helgaker}},\ }\href
  {\doibase 10.1063/1.4867005} {\bibfield  {journal} {\bibinfo  {journal} {The
  Journal of Chemical Physics}\ }\textbf {\bibinfo {volume} {140}},\ \bibinfo
  {pages} {18A518} (\bibinfo {year} {2014})},\ \Eprint
  {http://arxiv.org/abs/https://doi.org/10.1063/1.4867005}
  {https://doi.org/10.1063/1.4867005} \BibitemShut {NoStop}%
\bibitem [{\citenamefont {Lieb}(1983)}]{DFT_math}%
  \BibitemOpen
  \bibfield  {author} {\bibinfo {author} {\bibfnamefont {E.~H.}\ \bibnamefont
  {Lieb}},\ }\href {\doibase 10.1002/qua.560240302} {\bibfield  {journal}
  {\bibinfo  {journal} {International Journal of Quantum Chemistry}\ }\textbf
  {\bibinfo {volume} {24}},\ \bibinfo {pages} {243} (\bibinfo {year}
  {1983})}\BibitemShut {NoStop}%
\bibitem [{\citenamefont {Chayes}\ \emph {et~al.}(1985)\citenamefont {Chayes},
  \citenamefont {Chayes},\ and\ \citenamefont {Ruskai}}]{Chayes_Ruskai}%
  \BibitemOpen
  \bibfield  {author} {\bibinfo {author} {\bibfnamefont {J.~T.}\ \bibnamefont
  {Chayes}}, \bibinfo {author} {\bibfnamefont {L.}~\bibnamefont {Chayes}}, \
  and\ \bibinfo {author} {\bibfnamefont {M.~B.}\ \bibnamefont {Ruskai}},\
  }\href {\doibase 10.1007/BF01010474} {\bibfield  {journal} {\bibinfo
  {journal} {Journal of Statistical Physics}\ }\textbf {\bibinfo {volume}
  {38}},\ \bibinfo {pages} {497} (\bibinfo {year} {1985})}\BibitemShut
  {NoStop}%
\bibitem [{\citenamefont {Ruggenthaler}\ \emph {et~al.}(2015)\citenamefont
  {Ruggenthaler}, \citenamefont {Penz},\ and\ \citenamefont
  {Van~Leeuwen}}]{existence_TDDFT}%
  \BibitemOpen
  \bibfield  {author} {\bibinfo {author} {\bibfnamefont {M.}~\bibnamefont
  {Ruggenthaler}}, \bibinfo {author} {\bibfnamefont {M.}~\bibnamefont {Penz}},
  \ and\ \bibinfo {author} {\bibfnamefont {R.}~\bibnamefont {Van~Leeuwen}},\
  }\href@noop {} {\bibfield  {journal} {\bibinfo  {journal} {Journal of
  Physics: Condensed Matter}\ }\textbf {\bibinfo {volume} {27}},\ \bibinfo
  {pages} {203202} (\bibinfo {year} {2015})}\BibitemShut {NoStop}%
\bibitem [{\citenamefont {Gilbert}(1975)}]{Gilbert:75}%
  \BibitemOpen
  \bibfield  {author} {\bibinfo {author} {\bibfnamefont {T.~L.}\ \bibnamefont
  {Gilbert}},\ }\href@noop {} {\bibfield  {journal} {\bibinfo  {journal} {Phys.
  Rev. B}\ }\textbf {\bibinfo {volume} {12}},\ \bibinfo {pages} {2111}
  (\bibinfo {year} {1975})}\BibitemShut {NoStop}%
\bibitem [{\citenamefont {Schindlmayr}\ and\ \citenamefont
  {Godby}(1995)}]{v-representability_gamma_lattice}%
  \BibitemOpen
  \bibfield  {author} {\bibinfo {author} {\bibfnamefont {A.}~\bibnamefont
  {Schindlmayr}}\ and\ \bibinfo {author} {\bibfnamefont {R.~W.}\ \bibnamefont
  {Godby}},\ }\href {\doibase 10.1103/PhysRevB.51.10427} {\bibfield  {journal}
  {\bibinfo  {journal} {Phys. Rev. B}\ }\textbf {\bibinfo {volume} {51}},\
  \bibinfo {pages} {10427} (\bibinfo {year} {1995})}\BibitemShut {NoStop}%
\bibitem [{\citenamefont {L\'opez-Sandoval}\ and\ \citenamefont
  {Pastor}(2002)}]{Pastor}%
  \BibitemOpen
  \bibfield  {author} {\bibinfo {author} {\bibfnamefont {R.}~\bibnamefont
  {L\'opez-Sandoval}}\ and\ \bibinfo {author} {\bibfnamefont {G.~M.}\
  \bibnamefont {Pastor}},\ }\href {\doibase 10.1103/PhysRevB.66.155118}
  {\bibfield  {journal} {\bibinfo  {journal} {Phys. Rev. B}\ }\textbf {\bibinfo
  {volume} {66}},\ \bibinfo {pages} {155118} (\bibinfo {year}
  {2002})}\BibitemShut {NoStop}%
\bibitem [{\citenamefont {Kamil}\ \emph {et~al.}(2016)\citenamefont {Kamil},
  \citenamefont {Schade}, \citenamefont {Pruschke},\ and\ \citenamefont
  {Bl\"ochl}}]{Mueller_Hubbard}%
  \BibitemOpen
  \bibfield  {author} {\bibinfo {author} {\bibfnamefont {E.}~\bibnamefont
  {Kamil}}, \bibinfo {author} {\bibfnamefont {R.}~\bibnamefont {Schade}},
  \bibinfo {author} {\bibfnamefont {T.}~\bibnamefont {Pruschke}}, \ and\
  \bibinfo {author} {\bibfnamefont {P.~E.}\ \bibnamefont {Bl\"ochl}},\ }\href
  {\doibase 10.1103/PhysRevB.93.085141} {\bibfield  {journal} {\bibinfo
  {journal} {Phys. Rev. B}\ }\textbf {\bibinfo {volume} {93}},\ \bibinfo
  {pages} {085141} (\bibinfo {year} {2016})}\BibitemShut {NoStop}%
\bibitem [{\citenamefont {Vignale}\ and\ \citenamefont
  {Rasolt}(1987)}]{VignaleRasolt:87}%
  \BibitemOpen
  \bibfield  {author} {\bibinfo {author} {\bibfnamefont {G.}~\bibnamefont
  {Vignale}}\ and\ \bibinfo {author} {\bibfnamefont {M.}~\bibnamefont
  {Rasolt}},\ }\href {\doibase 10.1103/PhysRevLett.59.2360} {\bibfield
  {journal} {\bibinfo  {journal} {Phys. Rev. Lett.}\ }\textbf {\bibinfo
  {volume} {59}},\ \bibinfo {pages} {2360} (\bibinfo {year}
  {1987})}\BibitemShut {NoStop}%
\bibitem [{\citenamefont {Diener}(1991)}]{Diener:91}%
  \BibitemOpen
  \bibfield  {author} {\bibinfo {author} {\bibfnamefont {G.}~\bibnamefont
  {Diener}},\ }\href {http://stacks.iop.org/0953-8984/3/i=47/a=014} {\bibfield
  {journal} {\bibinfo  {journal} {Journal of Physics: Condensed Matter}\
  }\textbf {\bibinfo {volume} {3}},\ \bibinfo {pages} {9417} (\bibinfo {year}
  {1991})}\BibitemShut {NoStop}%
\bibitem [{\citenamefont {Vignale}(2004)}]{Vignale:04}%
  \BibitemOpen
  \bibfield  {author} {\bibinfo {author} {\bibfnamefont {G.}~\bibnamefont
  {Vignale}},\ }\href {\doibase 10.1103/PhysRevB.70.201102} {\bibfield
  {journal} {\bibinfo  {journal} {Phys. Rev. B}\ }\textbf {\bibinfo {volume}
  {70}},\ \bibinfo {pages} {201102} (\bibinfo {year} {2004})}\BibitemShut
  {NoStop}%
\bibitem [{\citenamefont {Tokatly}(2011)}]{TDCDT_lattice_proof}%
  \BibitemOpen
  \bibfield  {author} {\bibinfo {author} {\bibfnamefont {I.~V.}\ \bibnamefont
  {Tokatly}},\ }\href {\doibase 10.1103/PhysRevB.83.035127} {\bibfield
  {journal} {\bibinfo  {journal} {Phys. Rev. B}\ }\textbf {\bibinfo {volume}
  {83}},\ \bibinfo {pages} {035127} (\bibinfo {year} {2011})}\BibitemShut
  {NoStop}%
\bibitem [{\citenamefont {Ullrich}(2012)}]{Ullrich:12}%
  \BibitemOpen
  \bibfield  {author} {\bibinfo {author} {\bibfnamefont {C.~A.}\ \bibnamefont
  {Ullrich}},\ }\href@noop {} {\emph {\bibinfo {title} {{T}ime-{D}ependent
  {D}ensity-{F}unctional {T}heory: {C}oncepts and {A}pplications}}},\ Oxford
  Graduate Texts\ (\bibinfo  {publisher} {Oxford University Press},\ \bibinfo
  {address} {Oxford},\ \bibinfo {year} {2012})\BibitemShut {NoStop}%
\bibitem [{\citenamefont {van Leeuwen}\ and\ \citenamefont
  {Baerends}(1994)}]{RVL_inversion}%
  \BibitemOpen
  \bibfield  {author} {\bibinfo {author} {\bibfnamefont {R.}~\bibnamefont {van
  Leeuwen}}\ and\ \bibinfo {author} {\bibfnamefont {E.~J.}\ \bibnamefont
  {Baerends}},\ }\href {\doibase 10.1103/PhysRevA.49.2421} {\bibfield
  {journal} {\bibinfo  {journal} {Phys. Rev. A}\ }\textbf {\bibinfo {volume}
  {49}},\ \bibinfo {pages} {2421} (\bibinfo {year} {1994})}\BibitemShut
  {NoStop}%
\bibitem [{\citenamefont {Wu}\ and\ \citenamefont
  {Yang}(2003)}]{direct_optimization_inversion}%
  \BibitemOpen
  \bibfield  {author} {\bibinfo {author} {\bibfnamefont {Q.}~\bibnamefont
  {Wu}}\ and\ \bibinfo {author} {\bibfnamefont {W.}~\bibnamefont {Yang}},\
  }\href {\doibase 10.1063/1.1535422} {\bibfield  {journal} {\bibinfo
  {journal} {The Journal of Chemical Physics}\ }\textbf {\bibinfo {volume}
  {118}},\ \bibinfo {pages} {2498} (\bibinfo {year} {2003})},\ \Eprint
  {http://arxiv.org/abs/http://aip.scitation.org/doi/pdf/10.1063/1.1535422}
  {http://aip.scitation.org/doi/pdf/10.1063/1.1535422} \BibitemShut {NoStop}%
\bibitem [{\citenamefont {Jensen}\ and\ \citenamefont
  {Wasserman}(2018)}]{Inversion_review_Wasserman}%
  \BibitemOpen
  \bibfield  {author} {\bibinfo {author} {\bibfnamefont {D.~S.}\ \bibnamefont
  {Jensen}}\ and\ \bibinfo {author} {\bibfnamefont {A.}~\bibnamefont
  {Wasserman}},\ }\href {\doibase 10.1002/qua.25425} {\bibfield  {journal}
  {\bibinfo  {journal} {International Journal of Quantum Chemistry}\ }\textbf
  {\bibinfo {volume} {118}},\ \bibinfo {pages} {e25425} (\bibinfo {year}
  {2018})},\ \bibinfo {note} {e25425}\BibitemShut {NoStop}%
\bibitem [{\citenamefont {Ryabinkin}\ \emph {et~al.}(2015)\citenamefont
  {Ryabinkin}, \citenamefont {Kohut},\ and\ \citenamefont
  {Staroverov}}]{Staroverov_inversion}%
  \BibitemOpen
  \bibfield  {author} {\bibinfo {author} {\bibfnamefont {I.~G.}\ \bibnamefont
  {Ryabinkin}}, \bibinfo {author} {\bibfnamefont {S.~V.}\ \bibnamefont
  {Kohut}}, \ and\ \bibinfo {author} {\bibfnamefont {V.~N.}\ \bibnamefont
  {Staroverov}},\ }\href {\doibase 10.1103/PhysRevLett.115.083001} {\bibfield
  {journal} {\bibinfo  {journal} {Phys. Rev. Lett.}\ }\textbf {\bibinfo
  {volume} {115}},\ \bibinfo {pages} {083001} (\bibinfo {year}
  {2015})}\BibitemShut {NoStop}%
\bibitem [{\citenamefont {Hubig}\ \emph {et~al.}(2015)\citenamefont {Hubig},
  \citenamefont {McCulloch}, \citenamefont {Schollw\"ock},\ and\ \citenamefont
  {Wolf}}]{hubig15}%
  \BibitemOpen
  \bibfield  {author} {\bibinfo {author} {\bibfnamefont {C.}~\bibnamefont
  {Hubig}}, \bibinfo {author} {\bibfnamefont {I.~P.}\ \bibnamefont
  {McCulloch}}, \bibinfo {author} {\bibfnamefont {U.}~\bibnamefont
  {Schollw\"ock}}, \ and\ \bibinfo {author} {\bibfnamefont {F.~A.}\
  \bibnamefont {Wolf}},\ }\href {\doibase 10.1103/PhysRevB.91.155115}
  {\bibfield  {journal} {\bibinfo  {journal} {Phys. Rev. B}\ }\textbf {\bibinfo
  {volume} {91}},\ \bibinfo {pages} {155115} (\bibinfo {year}
  {2015})}\BibitemShut {NoStop}%
\bibitem [{\citenamefont {Hubig}(2017)}]{hubig17_2}%
  \BibitemOpen
  \bibfield  {author} {\bibinfo {author} {\bibfnamefont {C.}~\bibnamefont
  {Hubig}},\ }\emph {\bibinfo {title} {Symmetry-Protected Tensor Networks}},\
  \href {https://edoc.ub.uni-muenchen.de/21348/} {Ph.D. thesis},\ \bibinfo
  {school} {LMU München} (\bibinfo {year} {2017})\BibitemShut {NoStop}%
\bibitem [{\citenamefont {Ruggenthaler}\ and\ \citenamefont
  {Bauer}(2009)}]{local_force_Mx}%
  \BibitemOpen
  \bibfield  {author} {\bibinfo {author} {\bibfnamefont {M.}~\bibnamefont
  {Ruggenthaler}}\ and\ \bibinfo {author} {\bibfnamefont {D.}~\bibnamefont
  {Bauer}},\ }\href {\doibase 10.1103/PhysRevA.80.052502} {\bibfield  {journal}
  {\bibinfo  {journal} {Phys. Rev. A}\ }\textbf {\bibinfo {volume} {80}},\
  \bibinfo {pages} {052502} (\bibinfo {year} {2009})}\BibitemShut {NoStop}%
\bibitem [{\citenamefont {Fuks}\ \emph {et~al.}(2016)\citenamefont {Fuks},
  \citenamefont {Nielsen}, \citenamefont {Ruggenthaler},\ and\ \citenamefont
  {Maitra}}]{fuks2016time}%
  \BibitemOpen
  \bibfield  {author} {\bibinfo {author} {\bibfnamefont {J.~I.}\ \bibnamefont
  {Fuks}}, \bibinfo {author} {\bibfnamefont {S.~E.}\ \bibnamefont {Nielsen}},
  \bibinfo {author} {\bibfnamefont {M.}~\bibnamefont {Ruggenthaler}}, \ and\
  \bibinfo {author} {\bibfnamefont {N.~T.}\ \bibnamefont {Maitra}},\
  }\href@noop {} {\bibfield  {journal} {\bibinfo  {journal} {Physical Chemistry
  Chemical Physics}\ }\textbf {\bibinfo {volume} {18}},\ \bibinfo {pages}
  {20976} (\bibinfo {year} {2016})}\BibitemShut {NoStop}%
\bibitem [{\citenamefont {Liao}\ \emph {et~al.}(2017)\citenamefont {Liao},
  \citenamefont {Ho}, \citenamefont {Rabitz},\ and\ \citenamefont
  {Chu}}]{liao2017time}%
  \BibitemOpen
  \bibfield  {author} {\bibinfo {author} {\bibfnamefont {S.-L.}\ \bibnamefont
  {Liao}}, \bibinfo {author} {\bibfnamefont {T.-S.}\ \bibnamefont {Ho}},
  \bibinfo {author} {\bibfnamefont {H.}~\bibnamefont {Rabitz}}, \ and\ \bibinfo
  {author} {\bibfnamefont {S.-I.}\ \bibnamefont {Chu}},\ }\href@noop {}
  {\bibfield  {journal} {\bibinfo  {journal} {Physical review letters}\
  }\textbf {\bibinfo {volume} {118}},\ \bibinfo {pages} {243001} (\bibinfo
  {year} {2017})}\BibitemShut {NoStop}%
\bibitem [{Note1()}]{Note1}%
  \BibitemOpen
  \bibinfo {note} {We want to point out, that for larger values of $U$ we
  encountered some convergence issues in the 4-site case. The reason being that
  while the density-matrices are not homogeneous, the density is, which causes
  some problems in the iteration scheme, where we divide by the density
  difference in each iteration. This problem, however, can be potentially
  overcome by using different update equations.}\BibitemShut {Stop}%
\bibitem [{\citenamefont {Griesemer}\ and\ \citenamefont
  {Hantsch}(2012)}]{griesemer2012unique}%
  \BibitemOpen
  \bibfield  {author} {\bibinfo {author} {\bibfnamefont {M.}~\bibnamefont
  {Griesemer}}\ and\ \bibinfo {author} {\bibfnamefont {F.}~\bibnamefont
  {Hantsch}},\ }\href@noop {} {\bibfield  {journal} {\bibinfo  {journal}
  {Archive for Rational Mechanics and Analysis}\ }\textbf {\bibinfo {volume}
  {203}},\ \bibinfo {pages} {883} (\bibinfo {year} {2012})}\BibitemShut
  {NoStop}%
\end{thebibliography}%

\end{document}